%% file: main.tex
\documentclass{article}

\usepackage[english]{babel}
\usepackage[a4paper,top=2cm,bottom=2cm,left=3cm,right=3cm,marginparwidth=1.75cm]{geometry}

\usepackage{natbib}
\usepackage{amsmath}
\usepackage{amssymb}
\usepackage{amsthm}
\usepackage{mathtools}\mathtoolsset{showonlyrefs=true}
\usepackage{graphicx}
\usepackage{bm}
\usepackage{xcolor}
\usepackage[colorlinks=true, allcolors=blue]{hyperref}
\usepackage{longtable}
\usepackage{booktabs}
\usepackage{multirow}
\usepackage{float}
\usepackage{caption}
\usepackage{subcaption}
\usepackage{tabularx} 
\usepackage{array}
\usepackage{algpseudocode}

\usepackage{dsfont}
\usepackage[ruled,vlined]{algorithm2e}

\usepackage{color}
\usepackage[normalem]{ulem}

\newtheorem{lemma}{Lemma}

\providecommand{\keywords}[1]{\textbf{\textit{Keywords---}} #1}

\newcommand{\EE}{\mathbb{E}} 
\newcommand{\RR}{\mathbb{R}} 
\newcommand{\NN}{\mathbb{N}} 
\newcommand{\T}{\mathrm{T}} 

\newcommand{\eps}{\epsilon}
\newcommand{\trace}{\mathrm{tr}}

\newcommand{\TT}[1]{\mathcal{T}_{#1}} 
\newcommand{\sLp}[1]{\mathcal{L}^{2}\left(#1\right)} 
\newcommand{\HH}{\mathcal{H}} 

\newcommand{\hatK}{\widehat{\mathbf{K}}}
\newcommand{\tildeK}{\widetilde{\mathbf{K}}}
\newcommand{\Smoother}{\mathbf{S}}
\newcommand{\B}{\mathbf{B}}
\newcommand{\Penalty}{\mathbf{P}}
\newcommand{\Dmat}[1]{\mathbf{D}_{#1}}
\newcommand{\BTB}[1]{(\B^{\T} \B)^{#1}}
\newcommand{\G}{\mathbf{G}}
\newcommand{\U}{\mathbf{U}}
\newcommand{\A}{\mathbf{A}}
\newcommand{\Sigmas}{\mathbf{\Sigma}}
\newcommand{\s}{\mathbf{s}}

\newcommand{\TTheta}{\mathbf{\Theta}}
\newcommand{\Q}{\mathbf{Q}}
\newcommand{\V}{\mathbf{V}}
\newcommand{\Bder}{\textbf{B}_{\text{der}}}
\newcommand{\Yder}{\mathbf{Y}_i^{d}}
\newcommand{\PPhi}{\mathbf{\Phi}}
\newcommand{\xxi}{\bm{\xi}}
\newcommand{\ttheta}{\bm{\theta}}

\newcommand{\Xp}[1]{X^{[#1]}} 
\newcommand{\Xip}[1]{X_i^{[#1]}} 
\newcommand{\pointt}{\mathbf{t}}
\newcommand{\pointTT}{\bm{\mathcal{T}}}
\newcommand{\XXi}{\bm{\Xi}}
\newcommand{\cbold}{\mathbf{c}}

\newcommand{\diff}{\partial^d}

\DeclareMathOperator{\Var}{Var}
\DeclareMathOperator{\Cov}{Cov}

\newcommand\restr[2]{{ %
  \left.\kern-\nulldelimiterspace  %
  #1  %
  \vphantom{\big|}  %
  \right|_{#2}  %
}}

\title{Covariance estimation for derivatives of functional data using an additive penalty in P-splines}
\author{%
Yueyun Zhu\thanks{School of Mathematical and Statistical Sciences, University of Galway, Ireland \href{mailto:yueyun.zhu@universityofgalway.ie}{yueyun.zhu@universityofgalway.ie}}
\and
Steven Golovkine\thanks{Department of Mathematics and Statistics, Université Laval, Canada \href{mailto:steven.golovkine@mat.ulaval.ca}{steven.golovkine@mat.ulaval.ca}}
\and
Norma Bargary\thanks{MACSI, Department of Mathematics and Statistics, University of Limerick, Ireland \href{mailto:norma.bargary@ul.ie}{norma.bargary@ul.ie}}
\and
Andrew J. Simpkin\thanks{School of Mathematical and Statistical Sciences, University of Galway, Ireland \href{mailto:andrew.simpkin@universityofgalway.ie}{andrew.simpkin@universityofgalway.ie}}
}
\date{\today}

\begin{document}
\maketitle

\begin{abstract}

P-splines provide a flexible and computationally efficient smoothing framework and are commonly used for derivative estimation in functional data. Including an additive penalty term in P-splines has been shown to improve estimates of derivatives. We propose a method which incorporates the fast covariance estimation (FACE) algorithm with an additive penalty in P-splines. The proposed method is used to estimate derivatives of covariance for functional data, which play an important role in derivative-based functional principal component analysis (FPCA). Following this, we provide an algorithm for estimating the eigenfunctions and their corresponding scores in derivative-based FPCA. For comparison, we evaluate our algorithm against an existing function \texttt{FPCAder()} in simulation. In addition, we extend the algorithm to multivariate cases, referred to as derivative multivariate functional principal component analysis (DMFPCA). DMFPCA is applied to joint angles in human movement data, where the derivative-based scores demonstrate strong performance in distinguishing locomotion tasks.
\end{abstract}


\input{main/introduction}

\input{main/FACE_with_additive_penalty}

\input{main/derivative-based_FPCA}

\input{main/simulation}
 
\input{main/application}

\input{main/discussion}

\input{main/appendix}



\section*{Acknowledgment}

This work was supported by the Functional data Analysis for Sensor Technologies (FAST) project, funded by Research Ireland grant 19/FFP/7002.

\bibliographystyle{apalike}
\bibliography{./biblio.bib}

\end{document}

%% file: main/introduction.tex
\section{Introduction} 
\label{sec:introduction}

With the development of wearable monitoring devices, sensors, and neuroimaging technologies, increasingly large and complex datasets are being recorded. Functional data analysis (FDA) has become important in this context, with early foundational work developed by \cite{rao1958some,ramsay1982data} and \cite{ramsay1991some}. 

Functional principal component analysis (FPCA) is a powerful tool in FDA. Various FPCA methods have been proposed and applied to analyze sensor and neuroimaging data for feature extraction, dimension reduction, regression and classification. For instance, \cite{di2009multilevel} introduced multilevel FPCA which was applied to the electroencephalographic (EEG) signals from the Sleep Heart Health Study to extract core intra- and inter-subject geometric components. \cite{happ2018multivariate} introduced multivariate FPCA which was applied to images from Alzheimer’s Disease Neuroimaging Initiative study to capture important sources of variation in the neuroimaging data. More applications of FPCA to biomechanics and human movement data can be found in \cite{coffey2011common, gunning2024functional, gunning2025analysing}.

Covariance estimation plays an important role in FPCA \citep{ramsay2002applied, Ramsay2005FDA, yao2005functional, wang2016functional, gertheiss2024functional}. The orthonormal eigenfunctions obtained by decomposing the covariance of functional trajectories constitute the functional principal components (FPCs), which capture the dominant modes of variation of the underlying function. Note that FPCA is often enhanced by the use of smoothing \citep{silverman1996smoothed}. Covariance smoothing is essential in FPCA because it reduces the measurement error and enables accurate estimation of the true FPCs and their eigenvalues. There are three main approaches to estimating the smooth covariance, namely ``smooth then estimate",``estimate then smooth" and ``low-rank smoothing". The ``smooth then estimate" first pre-smooths each trajectory independently and then estimates the covariance from the smoothed data, which can lead to substantial bias as the pre-smoothing procedure fails to capture the overall dependence structure. The ``low-rank smoothing" estimates the covariance by representing it as a low-dimensional outer product of smooth basis-expanded latent functions, see \cite{mbaka2025estimating}. The ``estimate then smooth" first estimates the raw sample covariance and then smooths the resulting sample covariance using local linear smoothers \citep{fan1995local} or P-splines \citep{eilers2021practical}. The ``estimate then smooth" is applied in most of the literature, see \citet{staniswalis1998nonparametric, yao2003shrinkage, yao2005functional, yao2006penalized} for example. Recently, a fast P-spline method for bivariate smoothing is introduced by \cite{xiao2013fast}, where they proposed a ``sandwich smoother'' that improves computational efficiency to estimate the smooth covariance function. The sandwich smoother was shown to be much faster than local linear smoothers. Following this, \cite{xiao2016fast} introduced a fast covariance estimation (FACE) algorithm which has significant computational advantages over bivariate P-splines. \cite{li2020fast} extended the FACE algorithm to multivariate sparse functional data, where they estimated the auto-covariance functions for within-function correlations and cross-covariance functions for between-function correlations. \cite{cui2023fast} proposed a fast multilevel FPCA based on the FACE algorithm, which demonstrates higher computational efficiency than \cite{di2009multilevel}.

In many applications, the derivatives of functional trajectories are not directly observed but they are useful in capturing the underlying dynamics, such as velocity and acceleration. Derivative-based FPCA is therefore important, which captures the modes of variation of the unobserved derivatives (i.e., derivative-based FPCs) and provides a low-dimensional representation of these derivatives (i.e., derivative-based scores). Subsequently, the derivatives can be recovered by the Karhunen-Loève expansion. Previous work on derivative-based FPCA can be found in \cite{liu2009estimating, dai2018derivative, grithFunctionalPrincipalComponent2018}. In \cite{liu2009estimating}, the derivative-based FPCs were estimated by taking the derivative from both sides of the eigenequation of the conventional FPCA. In other words, the estimation of derivative-based FPCs was linked to the task of estimating a partial derivative of the sample covariance function, which was achieved by the local linear smoother. \cite{dai2018derivative} illustrated that the derivative-based FPCs estimated by \cite{liu2009estimating} were not orthogonal. To obtain the orthogonal derivative-based FPCs, they presented a new approach by estimating the derivative of the covariance function with respect to two dimensions (i.e., the mixed partial derivative). In this case, the derivative-based FPCs were obtained via the eigendecomposition of the mixed partial derivative of the covariance and therefore were orthogonal. Their method has been implemented in the \texttt{FPCAder()} function of the \textit{fdapace} package \citep{fdapace}. \cite{grithFunctionalPrincipalComponent2018} extended the derivative-based FPCA to a multi-dimensional domain, where they applied the eigendecomposition to a dual covariance matrix of the derivatives to obtain the derivative-based FPCs. However, all these methods rely on local linear smoothers to estimate the derivative of covariance function. 

In derivative-based FPCA, estimating the derivative of covariance function is a fundamental step. In this paper, we are interested in using P-splines to estimate the derivative of covariance. P-splines are computationally efficient and provide enough flexibility to capture non-linearities in functional data. More importantly, in one-dimensional cases, incorporating an extra penalty term in a P-spline allows for increased smoothing and improves the accuracy of derivative estimation for functional curves \citep{simpkin2013additive}. To estimate the derivative of the covariance function, we aim to extend FACE algorithm by incorporating an additive penalty term in P-splines. Analogous to one-dimensional cases, the additive penalty enables a stronger smooth which is particularly useful for derivative estimation of the covariance surface. It is further found that the derivative-based FPCs can be estimated without requiring the eigendecomposition of the derivative of covariance function. To estimate the derivative-based scores, we propose a method based on mixed model equations (MME) \citep{henderson1973sire}. Following this framework, we extend the derivative-based FPCA to multivariate cases, referred to as derivative multivariate functional principal component analysis (DMFPCA). 

The paper is organized as follows. In Section~\ref{sec:FACE with additive penalty}, we introduce a new approach which incorporates an additive penalty in P-splines with the FACE algorithm. Section~\ref{sec:derivative-based FPCA} shows the estimation of derivative-based FPCA, including the derivative-based FPCs, eigenvalues and scores; and the extension to DMFPCA. A simulation study to evaluate the proposed derivative-based FPCA and DMFPCA is provided in Section~\ref{sec:simulation}, where the \texttt{FPCAder()} function is used for comparison with our approach for univariate cases. Section~\ref{sec:application} shows an application of DMFPCA to human movement data (e.g., the joint angles of the knee, hip and ankle) collected by wearable sensors, where the scores demonstrate strong classification performance in distinguishing participants across different locomotion tasks. The paper concludes with a discussion in Section~\ref{sec:discussion}.


%% file: main/FACE_with_additive_penalty.tex
\section{Fast covariance estimation with an additive penalty in P-splines} 
\label{sec:FACE with additive penalty}

Let $X: \TT{} \rightarrow \RR$ be a random process defined in the Hilbert space $\HH$, such that $X \in \sLp{\TT{}}$ where $\TT{} \subset \RR$ denotes the observation domain. Assume $X(t)$ is $d$-times differentiable, with mean $\EE(X(t))=0$ and covariance $K(s,t)=\Cov(X(s), X(t))$. Suppose that $\left\{X_i, i=1,\dots,n\right\}$ is a collection of independent realizations of $X$. Let $Y_{ij}$ be the $j$th observation ($j=1,\dots,J$) of the process $X_i$ at a random time $T_{j} \in \TT{}$, $Y_{ij}=X_i(T_{j})+\eps_{ij}$, where $\eps_{ij}$ are normally distributed measurement errors with $\EE(\eps_{ij})$=0 and $\Var(\eps_{ij})=\sigma^2$. The covariance of the sample can be calculated by $\hat{K}(T_j, T_l)=n^{-1}\sum_{i=1}^{n}Y_{ij}Y_{il}$ at each pair of sampling points $(T_j, T_l)$. Let $\mathbf{Y_i}=(Y_{i1},\dots,Y_{iJ})^\T$ and $\mathbf{Y}=[\mathbf{Y_1}, \dots, \mathbf{Y_n}]$. We denote the covariance of the sample by $\hatK=n^{-1}\mathbf{Y}\mathbf{Y}^{\T}$.

\subsection{Covariance estimation}
\label{sec:FACE_covest}
Following \cite{xiao2016fast}, the FACE estimator of the covariance matrix can be represented as
\begin{equation}
\label{eq:FACE_K}
\tildeK=\Smoother \hatK \Smoother,
\end{equation}
where $\Smoother \in \RR^{J \times J}$ is a smoother symmetric matrix constructed using P-splines \citep{eilers1996flexible}, $\Smoother=\B(\B^{\T} \B + \lambda \Penalty)^{-1}\B^\T$. Here, we denote $\B \in \RR^{J \times c}$ as the matrix of the B-spline basis, such that $\B = [\B_1, \dots, \B_c]$, where $\B_k=(B_k(t_1), \dots, B_k(t_J))^{\T}$ are the B-spline basis functions for $k=1,\dots,c$. The matrix $\Penalty \in \RR^{c \times c}$ is a symmetric penalty matrix, such that $\Penalty = \Dmat{l}^{\T}\Dmat{l}$, where $\Dmat{l}$ is a difference operator of dimension $(c-l) \times c$ and $l$ is the penalty order. The scalar $\lambda \geq 0$ is the smoothing parameter, with larger values resulting in smoother estimates. 

\cite{simpkin2013additive} and \cite{hernandez2023derivative} introduced methods to include an additive penalty structure to P-splines to allow for extra smoothing, which should help prevent undersmoothing in derivative estimation. By incorporating an additive penalty term in P-splines, the smoother matrix is reconstructed by
\begin{equation}
	\label{eq:FACE_S}
	\Smoother=\B(\B^{\T} \B + \lambda_1 \Penalty_1 + \lambda_2 \Penalty_2)^{-1}\B^{\T},
\end{equation}
where $\Penalty_1=\Dmat{l_1}^{\T}\Dmat{l_1}$ and $\Penalty_2=\Dmat{l_2}^{\T}\Dmat{l_2}$ are penalty matrices penalized by smoothing parameters $\lambda_1$ and $\lambda_2$, respectively. When there is a single penalty in P-splines, \cite{xiao2016fast} showed that FACE starts with the decomposition $\BTB{-1/2}\Penalty\BTB{-1/2}=\U \text{diag}(\s) \U^{\T}$ where $\U$ is the matrix of eigenvectors and $\s$ is the vector of eigenvalues. When including an additive penalty in P-splines, we define $\lambda_{+}=\lambda_1 + \lambda_2$ to be an overall smoothing level and $w=\lambda_1/\lambda_{+} \in [0,1]$ to be a relative weight of $\lambda_1$. Then, the overall penalty matrix $\lambda_1 \Penalty_1 + \lambda_2 \Penalty_2$ can be represented by $\lambda_{+} (w\Penalty_1 + (1-w)\Penalty_2)$. For a given $w$, let $\U \text{diag}(\s) \U^{\T}$ be the decomposition of $\BTB{-1/2}(w\Penalty_1+ (1-w) \Penalty_2) \BTB{-1/2}$, such that
\begin{equation}
	\label{eq:FACE_eigdec}
	\BTB{-1/2} \left(w\Penalty_1 + (1-w)\Penalty_2 \right)\BTB{-1/2} = \U \text{diag}(\s) \U^{\T},
\end{equation}
where $\U \in \RR^{c \times c}$ is the matrix of eigenvectors and $\s$ is the vector of eigenvalues. By using the decomposition in \eqref{eq:FACE_eigdec}, the smoother matrix can be simplified in Lemma~\ref{lem:smoothermatrix}, with the proof provided in Appendix~\ref{sec:Lemma}.

\begin{lemma}
	\label{lem:smoothermatrix}
	Define $\A_0=(\B^{\T} \B)^{-1/2}\U \in \RR^{c \times c}$, $\A_s=\B \A_0 \in \RR^{J \times c}$, $\Sigmas_s = \left\{\mathbf{I}_c + \lambda_{+} \text{diag} (\s)\right\}^{-1} \in \RR^{c \times c}$. $\A_s$ has orthonormal columns, $\A_s^{\T}\A_s=\mathbf{I}_c$. The smoother matrix is rewritten as $\Smoother = \A_s \Sigmas_s \A_s^{\T}$.
\end{lemma}

Based on Lemma~\ref{lem:smoothermatrix}, replacing $\Smoother$ with $\A_s \Sigmas_s \A_s^{\T}$ into expression \eqref{eq:FACE_K} gives $\tildeK=\B \TTheta \B^T$, where
\begin{equation}
	\label{eq:FACE_Theta}
	\TTheta=\A_0 \Sigmas_s \A_0^{\T} \B^{\T} \hatK \B \A_0 \Sigmas_s \A_0^{\T}
\end{equation}
is a $c \times c$ symmetric and positive-definite matrix. More specifically, the $(s,t)$th entry of the covariance estimator is
\begin{equation}
	\label{eq:FACE_Kst}
	\widetilde{K}(s,t)=\B^{\T}(s)\TTheta \B(t).
\end{equation}

The covariance estimator is in `sandwich form'. The B-spline bases are uniquely determined once the knot, degree and the number of basis functions are given. Therefore, to obtain the covariance estimator, we only need to estimate $\TTheta$. We start with the selection of smoothing parameters. According to \citet[Prop.2]{xiao2016fast}, the smoothing parameters are estimated by minimizing a generalized cross-validation (GCV):
\begin{equation}
	\label{eq:FACE_gcv}
	\frac{\sum_{i=1}^{n}\| \mathbf{Y_i}-\Smoother \mathbf{Y_i} \|^2 }{(1-\trace(\Smoother)/J)^2}=
	\frac{\sum_{k=1}^{c}C_{kk}(\lambda_{+} s_k)^2/(1+\lambda_{+} s_k)^2-\| \mathbf{Y}_s  \|_F^{2} + \| \mathbf{Y} \|_F^{2} }{\left\{1-J^{-1}\sum_{k=1}^{c}(1+\lambda_{+} s_k)^{-1}\right\}^2},
\end{equation}
where $s_k$ is the $k$th element of $\s$, $\mathbf{Y}_s=\A_s^{\T}\mathbf{Y} \in \RR^{c \times n}$, $C_{kk}$ is the $k$th diagonal element of $\mathbf{Y}_s \mathbf{Y}_s^{\T}$, and $\| \cdot\|_{F}$ is the Frobenius norm. In practice, we perform a two-dimensional grid search of $(\lambda_{+}, w)$, building on a similar idea shown in \cite{li2020fast}. For a given pair $(\lambda_{+}, w)$, we calculate $\U$ and $\s$ based on the decomposition equation in \eqref{eq:FACE_eigdec}, then apply Lemma~\ref{lem:smoothermatrix} to construct $\A_s$ and finally calculate the GCV in \eqref{eq:FACE_gcv}. The optimal pair $(\lambda_{+}, w)$ is chosen by minimizing GCV over the two-dimensional grid, which is equivalent to re-parametrizing $\lambda_1=w\lambda_{+}$ and $\lambda_2=\lambda_{+}(1-w)$. After finding the optimal pair $(\lambda_{+}, w)$, we re-calculate $\U$, $\s$, $\A_0$ and $\Sigmas_s$ and replace them in \eqref{eq:FACE_Theta} to obtain $\TTheta$. The method of FACE with additive penalty is summarized in Algorithm~\ref{alg:FACE_with_additive_penalty}. 

\begin{algorithm}
	\caption{FACE with an additive penalty in P-splines}
	\label{alg:FACE_with_additive_penalty}
	\begin{algorithmic}
		\Require Sample $\left\{Y_{ij}\right\}_{1 \leq i \leq n, 1 \leq j \leq J}$, penalty orders $l_1$ and $l_2$, number of B-spline basis functions $c$, B-spline degree $q$.
		
		\item[\textbf{Construction:}] The B-spline matrix $\B$, the penalty matrices $\Penalty_1$ and $\Penalty_2$, the observation matrix $\mathbf{Y}$, the sample covariance matrix $\hatK$.
		
		\item[\textbf{Selection of smoothing parameters:}] For a given pair of $(\lambda_{+}, w)$,
		
		\State Step 1: Obtain $\U$ and $\s$ by solving \eqref{eq:FACE_eigdec}
		\State Step 2: Compute $\A_0$, $\A_s$, $\Sigmas_s$ by Lemma~\ref{lem:smoothermatrix} and $\mathbf{Y}_s = \A_s^{\T} \mathbf{Y}$
		\State Step 3: Evaluate the GCV in \eqref{eq:FACE_gcv} and identify the optimal pair $(\lambda_{+}, w)$ that minimizes GCV
		\State Step 4: Update $\U$, $\s$, $\A_0$, $\A_s$, $\Sigmas_s$ by using the selected $(\lambda_{+}, w)$
		
		\item[\textbf{Estimation of $\TTheta$}:] Calculate $\TTheta=\A_0 \Sigmas_s \A_0^{\T} \B^{\T} \hatK \B \A_0 \Sigmas_s \A_0^{\T}$ in \eqref{eq:FACE_Theta}.
		
		\item[\textbf{Output:}] The matrices $\A_0$, $\A_s$, $\Sigmas_s$, $\TTheta$.
	\end{algorithmic}
\end{algorithm}

\subsection{The derivative of covariance estimation}
\label{sec:FACE_dercovest}
Given that $X(t)$ is $d$-times differentiable, $K(s,t)$ is $d$-times differentiable in each argument. For simplicity, we denote 
\begin{equation}
	\label{eq:diff_notation}
	\diff X(t) \coloneqq \frac{\diff}{\partial t^d}X(t), \quad
	\diff \diff K(s,t) \coloneqq \frac{\partial^d}{\partial s^d}\frac{\partial^d}{\partial t^d} K(s,t), \quad
	d \in \NN
\end{equation}
where when $d=0$ in particular, we have $X(t) = \partial^0X(t)$ and $K(s,t) = \partial^0 \partial^0 K(s,t)$. Note that under the assumptions that $X(t)$ is mean centered, square integrable and $d$-times differentiable; $\diff \diff K(s,t)$ exists and is continuous; the derivative of covariance is equivalent to the covariance of derivatives, which is:
\begin{equation}
	\label{eq:diff_cov}
	\diff \diff K(s,t) = \frac{\partial^d}{\partial s^d}\frac{\partial^d}{\partial t^d} \Cov(X(s), X(t))
	=\EE \left( \diff X(s) \diff X(t) \right).
\end{equation}

Based on \citet{de1978practical} and \citet[Sec.2.5]{eilers2021practical}, B-splines have special derivative properties. For instance, assume $f(t) \in \HH$ can be represented by a linear combination of $c$ B-spline basis functions of degree $q$ defined over a sequence of evenly spaced knots. That is, $f(t)=\sum_{k=1}^{c}B_k(t;q)\theta_k$, or equivalently in the matrix form, $f(t)=\B^{\T}(t)\bm{\theta}$, where $\bm{\theta}=(\theta_1,\dots,\theta_c)^{\T}$ is a vector of coefficients and $\B(t)= \left\{B_k(t;q)\right\}_{k=1}^{c}$ is a vector of B-spline basis. The $d$th order derivative of $f$ can be obtained by $\diff f(t) = h^{-d} \left\{\B(t;q-d)\right\}^{\T} \Dmat{d} \bm{\theta}$, where $h$ is the distance between adjacent knots, $\Dmat{d} \in \RR^{(c-d) \times c}$ is a difference operator, $\B(t;q-d)= \left\{B_k(t;q-d)\right\}_{k=d+1}^{c} \in \RR^{(c-d) \times 1}$ is a vector of B-spline basis of degree $q-d$. This is a linear combination of the new B-spline basis functions of degree $q-d$ and the new coefficients after taking $d$th order difference of the original coefficient vector $\bm{\theta}$, divided by the constant $h^{d}$. On the other hand, the derivative of $f$ can be rewritten as $\diff f(t)=\Bder^{\T}(t)\bm{\theta}$, where $\Bder(t)=h^{-d}\Dmat{d}^{\T}\B(t;q-d)$. This shows the derivative is represented by a linear combination of the new basis $\Bder$ and the original vector of coefficients $\bm{\theta}$. Following this, the $d$th order derivative of the covariance estimator in \eqref{eq:FACE_Kst} can be expressed as
\begin{equation}
	\label{eq:FACE_diffK}
	\diff \diff \widetilde{K}(s,t) = \Bder^{\T}(s) \TTheta \Bder(t).
\end{equation}

The derivative of covariance retains the sandwich form, where $\Bder(t)$ is computed from $\B(t)$ and $\TTheta$ is the original coefficient matrix defined in \eqref{eq:FACE_Theta}. Algorithm~\ref{alg:FACE_with_additive_penalty} can therefore be directly applied to estimate the derivative of covariance. From the input, a derivative order $d$ will be needed. In the construction process, it is straightforward to have $\Bder$ once $\B$ is defined. The selection of smoothing parameters and the estimation of $\TTheta$ remain the same. By replacing $\Bder$ and $\TTheta$ in \eqref{eq:FACE_diffK}, the derivative of covariance is estimated.

To ensure the derivative of B-splines are continuous, we restrict the degree to satisfy $q \geq d+2$. In terms of the selection of penalty orders $l_1$ and $l_2$, \cite{simpkin2013additive} and \cite{hernandez2023derivative} compared several pairs of penalty order and found that the selections of penalty orders had little effect on the performance of derivative estimation (i.e., one-dimensional cases). In our study, we reach the same conclusion that the selection of penalty orders has minimal impact on the estimation of the derivative of covariance. We conduct a simulation with three pairs of penalty order $(l_1, l_2)=\left\{(1,2), (1,3), (2,3)\right\}$ and the results are provided in Section~\ref{sec:simulation_uni}.


%% file: main/derivative-based_FPCA.tex
\section{Derivative-based functional principal component analysis} 
\label{sec:derivative-based FPCA}

Covariance estimation is a fundamental step in FPCA \citep{ramsay2002applied, Ramsay2005FDA, yao2005functional}. With the approach introduced in Section~\ref{sec:FACE with additive penalty}, we can estimate the derivative of covariance, which allows conventional FPCA to be extended to derivative-based FPCA. While conventional FPCA focuses on variation in the functions themselves, derivative-based FPCA captures variation in their derivatives (e.g., velocities and accelerations), which is useful for analyzing underlying dynamics.

\subsection{Derivative-based eigenfunctions and eigenvalues}
\label{sec:derivative-based-eigfun}

The covariance operator $\Gamma: \HH \rightarrow \HH$ of $\diff X$ is defined as an integral operator with kernel $K_d$, where $K_d(s,t) = \Cov(\diff X(s), \diff X(t))$ is the covariance of derivatives. Assuming that $\Gamma$ is a compact positive operator on $\HH$, there exists an orthonormal basis $\left\{\phi_k\right\}_{k \geq 1}$ and a set of real numbers $\left\{\nu_{k}\right\}_{k \geq 1}$ such that $\nu_1 \geq \nu_2 \geq \dots \geq 0$, satisfying
\begin{equation}
	\Gamma \phi_k(\cdot) = \nu_k \phi_k(\cdot), \quad \nu_{k} \longrightarrow 0 \quad\text{as}\quad k \longrightarrow \infty.
\end{equation}
According to Mercer's theorem \citep{mercer1909functions}, the eigendecomposition of $K_d(s,t)$ can be expressed as $K_d(s,t) = \sum_{k=1}^{\infty}\nu_k\phi_k(s)\phi_k(t)$. As shown in Section~\ref{sec:FACE_dercovest}, the covariance of derivatives is equal to the derivative of covariance, i.e., $K_d(s,t) = \diff \diff K(s,t)$. Consequently, we can rewrite Mercer's expansion as
\begin{equation}
	\label{eq:eigen_equation}
	\int \diff \diff K(s,t) \phi_k(t) dt= \nu_k \phi_k(s).
\end{equation}

The smooth estimator of the derivative of the covariance, $\diff \diff \widetilde{K}(s,t)$, is continuous, symmetric and positive semi-definite, see \eqref{eq:FACE_diffK}. As $\diff \diff \widetilde{K}$ is expanded in $\Bder$, its eigenfunctions $\hat{\phi}_k$ should lie in the same functional space, which can be represented by $\hat{\phi}_k(t)=\Bder^{\T}(t)\hat{\bm{\alpha}}_k$, where $\hat{\bm{\alpha}}_k \in \RR^{c \times 1}$ is a coefficient vector associated to the $k$th derivative-based eigenfunction. Moreover, by replacing $\diff \diff K(s,t)$ with its smooth estimator, \eqref{eq:eigen_equation} becomes 
\begin{equation}
	\Bder^{\T}(s) \TTheta \left\{\int \Bder(t)\Bder^{\T}(t)dt\right\} \hat{\bm{\alpha}}_k = \hat{\nu}_k \Bder^{\T}(s)\hat{\bm{\alpha}}_k.
\end{equation}

Define $\G=\int \Bder(t)\Bder^{\T}(t)dt$, which is a $c \times c$ positive definite matrix. The above equation is further simplified to $\TTheta \G \hat{\bm{\alpha}}_k = \hat{\nu}_k \hat{\bm{\alpha}}_k$. By multiplying both sides by $\G^{1/2}$, we obtain $\left(\G^{1/2}\TTheta \G^{1/2}\right) \G^{1/2}\hat{\bm{\alpha}}_k = \hat{\nu}_k \G^{1/2} \hat{\bm{\alpha}}_k$, where $\hat{\nu}_k$ are the eigenvalues of $\G^{1/2}\TTheta \G^{1/2}$ and $\G^{1/2}\hat{\bm{\alpha}}_k$ are the corresponding eigenvectors. Let $\Q=[\Q_1, \dots, \Q_c]$ be an orthonormal matrix of eigenvectors, such that $\G^{1/2}\TTheta \G^{1/2}=\Q \hat{\V} \Q$, where $\hat{\V} =\text{diag}(\hat{\nu}_1, \dots, \hat{\nu}_c)$ is a diagonal matrix with the first $c$ eigenvalues of $\diff \diff \widetilde{K}$. It is straightforward to obtain $\hat{\bm{\alpha}}_k = \G^{-1/2}\Q_k$. Therefore, we finally have
\begin{equation}
	\label{eq:deriv_based_eigfun}
	\hat{\phi}_k(t) = \Bder^{\T}(t)\G^{-1/2}\Q_k, \quad k=1,\dots,c.
\end{equation}

Note that \eqref{eq:deriv_based_eigfun} provides a method to estimate the derivative-based eigenfunctions $\phi_k$ without requiring eigendecomposition of the derivative of covariance $\diff \diff \widetilde{K}$. The procedure for estimating the derivative-based eigencomponents is provided in Algorithm~\ref{alg:deriv_based_eigencomponents}.

\subsection{Derivative-based scores}

The best linear unbiased prediction (BLUP) is commonly used to estimate functional principal component scores \citep{yao2005functional, dai2018derivative}, and this requires the inversion of matrices that are of dimension equal to the number of observations per curve. \cite{cui2023fast} proposed a new method for multilevel FPCA based on a mixed model equation (MME), which reduces the computational complexity. We aim to extend the MME method to estimate the derivative-based scores.

Given that $\diff X(t)$ exists and has a finite second moment in the Hilbert space $\HH$, it can be represented by a Karhunen-Loève expansion
\begin{equation}
	\label{eq:KL_derX}
	\diff X(t) = \sum_{k=1}^{\infty}\xi_{k}\phi_k(t), \quad t \in \TT{},
\end{equation}
where $\phi_k(t)$ are the derivative-based eigenfunctions, $\xi_{k}$ are the derivative-based scores with $\xi_k \sim \mathcal{N}(0, \nu_k)$ and $\nu_k$ are the derivative-based eigenvalues. In practice, we use a truncated form of the Karhunen-Loève representation $\diff X(t) = \sum_{k=1}^{K}\xi_{k}\phi_k(t)$, where only the first $K$ eigenfunctions are selected. Let $\left\{\diff X_i, i=1,\dots,n\right\}$ be a collection of independent realizations of $\diff X$. Let $Y_{ij}^{d}$ be the $j$th observation of the process $\diff X_i$ at a random time $T_{j} \in \TT{}$,
\begin{equation}
	\label{eq:KL_derY}
	Y_{ij}^{d}= \sum_{k=1}^{K}\xi_{ik}\phi_k(t)+\eps_{ij}, \quad 
	\eps_{ij} \sim \mathcal{N}(0, \sigma^2).
\end{equation}

Define $\Yder=(Y_{i1}^{d}, \dots,Y_{iJ}^{d})^{\T}$, $\PPhi=[\phi_1, \dots, \phi_K]$, $\xxi_i=(\xi_{i1}, \dots, \xi_{iK})^{\T}$ and $\bm{\eps}_i=(\eps_{i1}, \dots, \eps_{iJ})^{\T}$. The model in \eqref{eq:KL_derY} is rewritten in a matrix form $\Yder=\PPhi\xxi_i+\bm{\eps}_i$. Note that $\xxi_i$ can be treated as a random effect. By using the MME \citep{henderson1973sire}, the derivative-based scores are represented as
\begin{equation}
	\label{eq:derv_based_scores}
	\xxi_i = (\PPhi^{\T}\PPhi + \sigma^2 \V^{-1})^{-1}\PPhi^{\T}\Yder,
\end{equation}
where $\V=\text{diag}(\nu_1, \dots, \nu_K)$ is a diagonal matrix with the first $K$ derivative-based eigenvalues. To estimate $\Yder$, we use the smoother matrix based on FACE with an additive penalty in P-splines. Let $\widetilde{\mathbf{Y}}_i$ be the smooth estimator of $\mathbf{Y}_i$, which is represented as $\widetilde{\mathbf{Y}}_i = \Smoother \mathbf{Y}_i$. By replacing $\Smoother$ with $\A_s \Sigmas_s \A_s^{\T}$ based on Lemma~\ref{lem:smoothermatrix}, $\widetilde{\mathbf{Y}}_i$ can be further expressed as $\widetilde{\mathbf{Y}}_i = \B \ttheta$, where $\ttheta = \A_0 \Sigmas_s \A_0^{\T}\B^{\T}\mathbf{Y}_i$. According to the special derivative properties of B-splines, the derivative of $\widetilde{\mathbf{Y}}_i$ can then be estimated by $\Yder= \Bder \ttheta$. For the estimation of $\sigma^2$, we have $\hat{\sigma}^2 = \int (\diff \diff \hat{K}(s,s) - \diff \diff \widetilde{K}(s,s))ds$, where $\diff \diff \hat{K}(s,s)$ and $\diff \diff \widetilde{K}(s,s)$ are the main diagonal elements of sample and smoothed covariance of the derivatives. The procedure of estimating the derivative-based scores is summarised in Algorithm~\ref{alg:deriv_based_eigencomponents}.

\begin{algorithm}
	\caption{Derivative-based eigencomponents and scores}
	\label{alg:deriv_based_eigencomponents}
	\begin{algorithmic}
		\Require Sample $\left\{Y_{ij}\right\}_{1 \leq i \leq n, 1 \leq j \leq J}$, penalty orders $l_1$ and $l_2$, derivative order $d$, number of B-spline basis functions $c$, B-spline degree $q$
		
		\item[\textbf{Construction:}] The B-spline matrix $\B$, the derivative B-spline matrix $\Bder$, the penalty matrices $\Penalty_1$ and $\Penalty_2$, the sample covariance matrix $\hatK$
		
		\item[\textbf{Estimation of $\TTheta$}:] Use FACE with an additive penalty in P-splines (Algorithm~\ref{alg:FACE_with_additive_penalty}) to obtain  $\TTheta$
		
		\item[\textbf{Estimation of derivative-based eigencomponents}:]
		\State Step 1: Calculate $\G=\int \Bder(t)\Bder^{\T}(t)dt$
		\State Step 2: Obtain $\Q$ and $\V$ by solving $\G^{1/2}\TTheta \G^{1/2}=\Q \V \Q$
		\State Step 3: Determine the smallest $K$ such that $\sum_{k=1}^{K}\nu_k/\sum_{k=1}^{c}\nu_k \geq 0.95$
		\State Step 4: Calculate the derivative-based eigenfunctions $\phi_k(t)$, $k=1,\dots, K$ based on \eqref{eq:deriv_based_eigfun} and select the associated derivative-based eigenvalues $\left\{\nu_1,\dots,\nu_K\right\}$
		
		\item[\textbf{Estimation of derivative-based scores}:]
		\State Step 1: Calculate $\ttheta = \A_0 \Sigmas_s \A_0^{\T}\B^{\T}\mathbf{Y}_i$, where $ \A_0$ and $\Sigmas_s$ are obtained from Algorithm~\ref{alg:FACE_with_additive_penalty}
		\State Step 2: Calculate $\Yder= \Bder \ttheta$ and construct $\mathbf{Y}^d=[\mathbf{Y}_1^d, \dots, \mathbf{Y}_n^d]$
		\State Step 3: Calculate $\diff \diff \hat{K}=n^{-1}\mathbf{Y}^d \left\{\mathbf{Y}^d\right\}^{\T}$ and $\diff \diff \tildeK = \Bder^{\T} \TTheta \Bder$
		\State Step 4: Calculate $\hat{\sigma}^2 = \int (\diff \diff \hat{K}(s,s) - \diff \diff \widetilde{K}(s,s))ds$
		\State Step 5: Calculate the derivative-based scores $\xxi_i$ based on \eqref{eq:derv_based_scores}
		
		\item[\textbf{Output:}] $\phi_k(t)$, $\left\{\nu_1, \dots, \nu_K\right\}$ and $\left\{\xi_{i1}, \dots, \xi_{iK}\right\}$
	\end{algorithmic}
\end{algorithm}

\subsection{Derivative multivariate functional principal component analysis}

\cite{happ2018multivariate} introduced an estimation strategy to calculate eigencomponents and scores of MFPCA based on their univariate counterparts. We extend the strategy to derivative multivariate functional principal component analysis (DMFPCA). 

We consider independent realizations of $X = (\Xp{1}, \dots, \Xp{P})^\T$, $P \geq 1$, a centered vector-valued stochastic process which consists of $P$ trajectories. Each $\Xp{p}: \TT{p} \rightarrow \RR$ is assumed to belong to  $\sLp{\TT{p}}$, with $\TT{p} \subset \RR$. Let $\pointTT \coloneqq \TT{1} \times \dots \times \TT{P}$ be the domain of $X$ and denote by $\pointt \coloneq (t_1, \ldots, t_P)$ an element of $\pointTT$. The process $X(\pointt) = (\Xp{1}(t_1), \dots, \Xp{P}(t_P))^\T$ is then defined on $\HH \coloneqq \sLp{\TT{1}} \times \dots \times \sLp{\TT{P}}$. Given that each $\diff X(\pointt)$ exists and has a finite second moment, it can be represented by a multivariate Karhunen-Loève expansion
\begin{equation}
	\label{eq:KL_multi_derX}
	\diff X(\pointt) = \sum_{k=1}^{\infty}\rho_{k}\psi_k(\pointt), \quad \pointt \in \pointTT,
\end{equation}
where $\rho_k$ are the derivative multivariate functional principal component scores (DMFPC-scores) and $\psi_k(\pointt) = (\psi_k^{[1]}(t_1), \dots, \psi_k^{[P]}(t_P))^{\T}$ are the derivative multivariate functional principal components (DMFPCs). 

In particular, a truncated form of \eqref{eq:KL_multi_derX} for the realizations of $\diff X$ can be written as $\diff X_i(\pointt) = \sum_{k=1}^{M}\rho_{ik}\psi_k(\pointt)$ where $M$ is the truncation number of DMFPCs and DMFPC-scores. We denote by $K_p$ the truncation number used for the $d$th derivative of $p$th univariate component, such that $\diff \Xip{p}(t)=\sum_{k=1}^{K_p}\xi_{ik}^{[p]}\phi_k^{[p]}(t)$. Note that the DMFPCA method requires only $M \ll K_+=\sum_{p=1}^{P}K_p$ scores, which achieves substantial dimension reduction and retains the essential variability of the derivatives. To estimate $\rho_k$ and $\psi_k$, we first construct a score matrix $\XXi \in \RR^{n \times K_+}$, where the $i$th row collects the $K_+$ derivative-based scores for all univariate components $p=1,\dots,P$, i.e.,
\begin{equation}
	\label{eq:DMFPCA_scorematrix}
	\begin{pmatrix*}
		\xi_{i1}^{[1]}, \dots, \xi_{iK_1}^{[1]}, \dots, \xi_{i1}^{[P]}, \dots, \xi_{iK_P}^{[P]}   
	\end{pmatrix*}.
\end{equation}

Define $\mathbf{Z}=(n-1)^{-1}\XXi^{\T}\XXi$. Let $v_k$ and $\cbold_k$ be the eigenvalues and eigenvectors of $\mathbf{Z}$. We define $\cbold_k=(\cbold_k^{[1]}, \dots, \cbold_k^{[P]})^{\T} \in \RR^{K_+ \times 1}$, with $\cbold_k^{[p]} \in \RR^{K_p \times 1}$ representing the $p$th block of $c_k$. We select the first $M$ eigenvectors and eigenvalues of $\mathbf{Z}$, where $M$ is the smallest number such that $\sum_{k=1}^{M}v_k / \sum_{k=1}^{K_+}v_k \geq 0.95$. Following \citet[Prop.5]{happ2018multivariate}, the elements of $\cbold_k^{[p]}$ can be used as the coefficients to link DMFPCs and DMFPC-scores to their univariate counterparts,
\begin{equation}
	\label{eq:DMFPCs_and_scores}
		\psi_k^{[p]}(t_p)=\sum_{l=1}^{K_p}\left(\cbold_k^{[p]}\right)_l \phi^{[p]}_{l}(t_p),\quad
		\rho_{ik}=\sum_{p=1}^{P}\sum_{l=1}^{K_p}\left(\cbold_k^{[p]}\right)_l\xi^{[p]}_{ik}, \quad k=1,\dots,M,
\end{equation}
where $\phi_l^{[p]}$ and $\xi_{ik}^{[p]}$ are the derivative-based eigenfunctions and scores for univariate data $\Xp{p}$ for $p=1,\dots,P$, and $\rho_{ik} \sim \mathcal{N}(0, v_k)$. The procedure of estimating the DMFPCA is provided in Algorithm~\ref{alg:DMFPCA}.

\begin{algorithm}
	\caption{DMFPCA}
	\label{alg:DMFPCA}
	\begin{algorithmic}
		\Require Sample $\left\{Y_{ij}^{[p]}\right\}_{1 \leq i \leq n, 1 \leq j \leq J, 1 \leq p \leq P}$, penalty orders $l_1$ and $l_2$, derivative order $d$, number of B-spline basis functions $c$, B-spline degree $q$
		
		\item[\textbf{For each $p \in [1,P]$}]: Estimate derivative-based eigenfunctions $\phi_k^{[p]}(t_p)$ and scores $\xi_{ik}^{[p]}$ using Algorithm~\ref{alg:deriv_based_eigencomponents}
		
		\item[\textbf{Construction}:] A score matrix $\XXi$ where $(\XXi)_{i\cdot}=(\xi_{i1}^{[1]}, \dots, \xi_{iK_1}^{[1]}, \dots, \xi_{i1}^{[P]}, \dots, \xi_{iK_P}^{[P]})$ and its covariance $\mathbf{Z}=(n-1)^{-1}\XXi^{\T}\XXi$
		
		\item[\textbf{Eigendecomposition}:] Obtain $\cbold_k$ and $v_k$ by solving $\mathbf{Z} \cbold_k=v_k\cbold_k$
		
		\item[\textbf{Selection of M}:] Determine the smallest $M$ such that $\sum_{k=1}^{M}v_k/\sum_{k=1}^{K_+}v_k \geq 0.95$
		
		\item[\textbf{Estimation of DMFPCs and DMFPC-scores:}] Calculate $\psi_k$ and $\rho_{ik}$ based on \eqref{eq:DMFPCs_and_scores}
		
		\item[\textbf{Output:}] $\psi_k(\pointt)$, $\left\{v_1, \dots, v_M\right\}$ and $\left\{\rho_{i1}, \dots, \rho_{iM}\right\}$
	\end{algorithmic}
\end{algorithm}


%% file: main/simulation.tex
\section{Simulation} 
\label{sec:simulation}

We simulate multivariate functional data with the following two univariate components:
\begin{equation}
	\begin{aligned}
		\Xp{1}(t) &= a + \frac{5}{ct + 10b\exp(-16t^2)},  \\
		\Xp{2}(t) &= a - \cos\left(\frac{ct}{4}(2t - \pi)\right) + 2\exp(-16bt^2),
	\end{aligned}
\end{equation}
where $a \sim \mathcal{N}(0,1)$, $b \sim \mathcal{N}(0.5,0.0196)$, $c \sim \mathcal{N}(3.75,0.49)$ and $\rho_{ab}=\rho_{ac}=\rho_{bc}=0.2$. We consider the most common case of the first derivative $d=1$ and use the notation $\partial X$ for simplicity. The derivatives of the two functions are:
\begin{equation}
\begin{aligned}
	\partial{} \Xp{1}(t) &= \frac{64bt\exp(-16t^2)-c/5}{(ct/5+2b\exp(-16t^2))^2}, \\
	\partial{} \Xp{2}(t) &= \left(ct-\frac{c\pi}{4}\right)\sin\left(\frac{ct}{4}(2t-\pi)\right)-64bt\exp(-16bt^2).\\
\end{aligned}
\end{equation}

\begin{figure}[H]
	\centering
	\begin{subfigure}{\textwidth}
		\centering
		\includegraphics[width=0.7\linewidth]{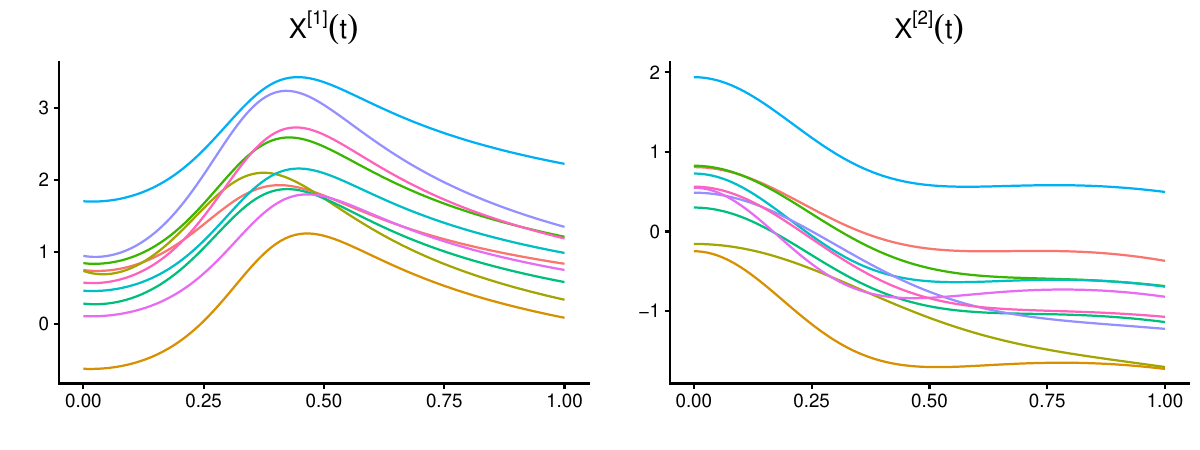}
	\end{subfigure}
	\\
	\begin{subfigure}{\textwidth}
		\centering
		\includegraphics[width=0.7\linewidth]{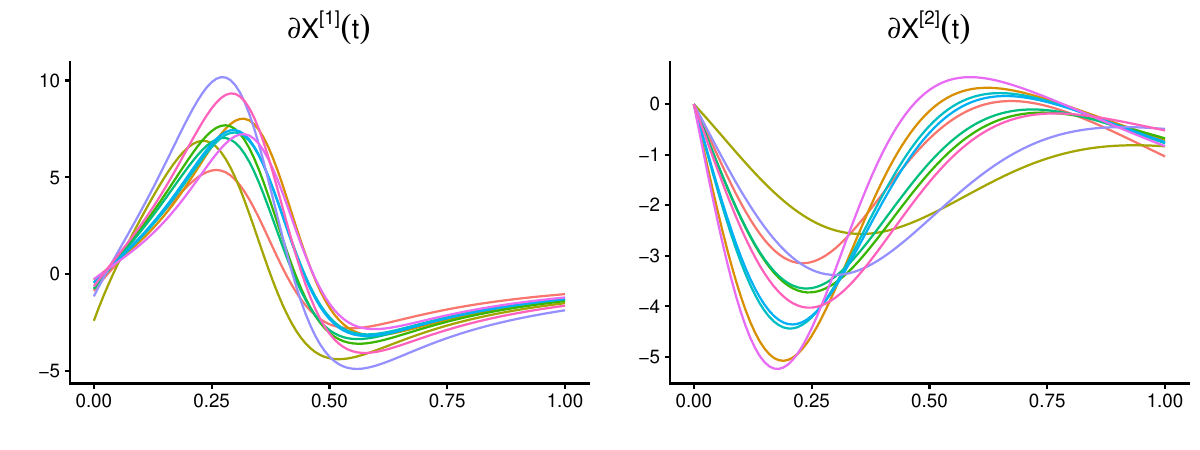}
	\end{subfigure}
	\caption{A sample of data simulated from the two functions (upper panel) and their corresponding true derivatives (lower panel).}
	\label{fig:sample_data}
\end{figure}

As an illustration, Figure~\ref{fig:sample_data} shows a sample of the data simulated from the two functions and their corresponding derivatives, when setting $t \in [0,1]$ with 101 equidistant observations. This indicates the trajectories of $\Xp{1}$ and $\Xp{2}$ are correlated and have some joint variation. For instance, the blue trajectory in $\Xp{1}(t)$ shows an increase over the first half of the domain, whereas the corresponding trajectory in $\Xp{2}$ exhibits a decrease over the same interval. Note that the data are non-centered. In the case of non-centered data, we use P-splines to estimate the mean and center the data at the initial step. In practice, the derivatives are usually unobserved. In this simulation, the derivatives are calculated straightforwardly since we have the functional forms of the data, which are treated as ground truth to evaluate the estimation methods.

We consider three settings: (1) densely observed data without measurement error; (2) densely observed data with measurement error $\sigma=0.5$; (3) sparsely observed data with measurement error $\sigma=0.5$. For each setting, we generate $n=100$ random trajectories for $p=1,2$, respectively, and run the simulation 500 times. For dense cases, data are observed on the same time grid $t_p \in [0,1]$ consisting of 101 equidistant observations. For the sparse case, data are observed on $t_p \in [0,1]$ while 40\%-50\% of the observations for each trajectory are missing completely at random. To handle sparse data, we reconstruct individual trajectories over the entire observation domain using the PACE approach of \cite{yao2005functional}.

\subsection{Univariate functional data}
\label{sec:simulation_uni}
We start with estimating the derivative-based eigencomponents and scores for univariate functional data. We apply Algorithm~\ref{alg:deriv_based_eigencomponents} to both $\Xp{1}$ and $\Xp{2}$ and refer to this method as `DFPCA'. For comparison, we use an existing R function \texttt{FPCAder()} \citep{dai2018derivative} in \textit{fdapace} package. In \cite{dai2018derivative}, a two-dimensional local polynomial smoothing method is employed to obtain the first-order derivative of the covariance function, and a BLUP method is used to estimate the derivative-based scores.

We evaluate accuracy using relative errors (RE) for eigenvalues, integrated square errors (ISE) for eigenfunctions, normalised mean squared errors (MSE) for scores and relative mean integrated square errors (RMISE) for derivatives. They are defined by: 
\begin{equation}
	\begin{aligned}
		\text{RE}(\nu_k^{[p]}, \widehat{\nu}_{k}^{[p]})&=\vert \nu_k^{[p]}-\widehat{\nu}_{k}^{[p]} \vert/\nu_k^{[p]},\\
		\text{ISE}(\phi_{k}^{[p]}, \widehat{\phi}_{k}^{[p]})&=\int \left(\phi_k^{[p]}(t_p)-\widehat{\phi}_k^{[p]}(t_p) \right)^2dt_p,\\
		\text{MSE}(\xi_k^{[p]}, \widehat{\xi}_k^{[p]})&=\frac{ \sum_{i=1}^{n}(\xi_{ik}^{[p]}-\widehat{\xi}_{ik}^{[p]})^2/n}{\Var(\xi_k^{[p]})},\\
		\text{RMISE}(\partial \Xp{p}, \widehat{\partial \Xp{p}})&=
		\frac{1}{n}\sum_{i=1}^{n}\frac{\int (\partial\Xp{p}_i(t_p)-\widehat{\partial \Xp{p}_i}(t_p))^2dt_p}
		{\int (\partial \Xp{p}_i(t_p))^2dt_p}.
	\end{aligned}
\end{equation}
Note that in simulation, $\nu_k^{[p]}$, $\phi_{k}^{[p]}$ and $\xi_k^{[p]}$ are obtained by applying R function \texttt{FPCA()} \citep{yao2005functional} to the true derivatives $\partial \Xp{p}$; $\widehat{\nu}_{k}^{[p]}$, $\widehat{\phi}_{k}^{[p]}$ and $\widehat{\xi}_k^{[p]}$ are estimated by applying DFPCA and FPCAder methods. After estimating the derivative-based eigenfunctions and scores, the Karhunen-Loève representation is used to construct the derivatives, i.e., $\widehat{\partial \Xp{p}}(t_p)= \sum_{k=1}^{K_p} \widehat{\xi}_k^{[p]}\widehat{\phi}_{k}^{[p]}(t_p)$, in comparison with the ground truth $\partial \Xp{p}$.

\begin{figure}[H]
	\centering
	\begin{subfigure}{\textwidth}
		\centering
		\includegraphics[width=\linewidth]{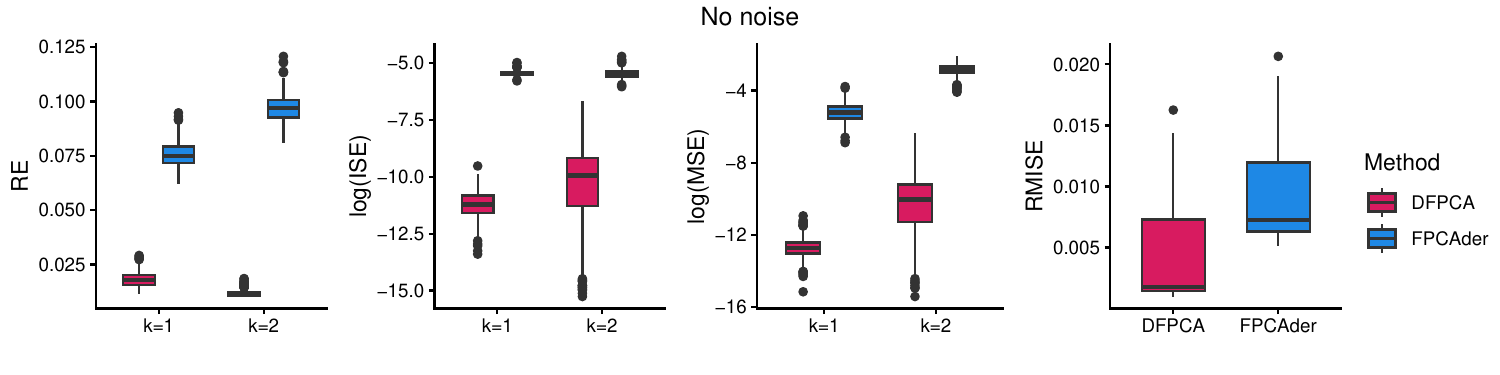}
	\end{subfigure}
	\\
	\begin{subfigure}{\textwidth}
		\centering
		\includegraphics[width=\linewidth]{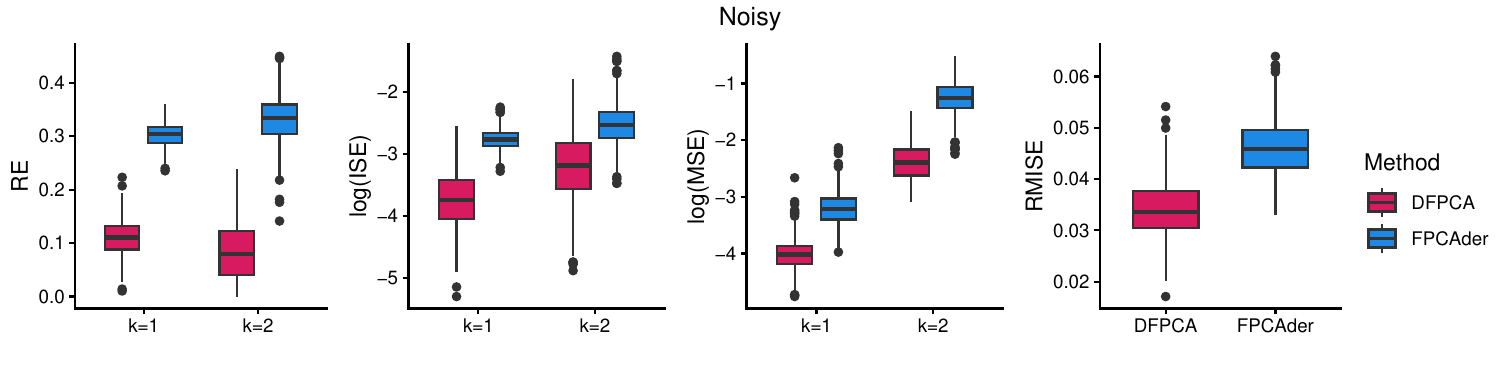}
	\end{subfigure}
	\\
	\begin{subfigure}{\textwidth}
		\centering
		\includegraphics[width=\linewidth]{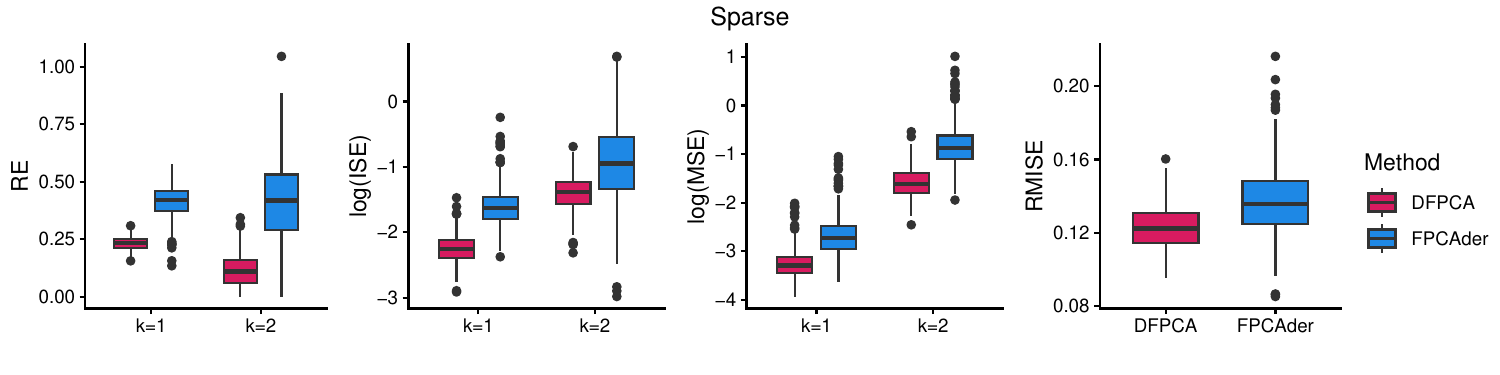}
	\end{subfigure}
	\caption{Simulation results for $\Xp{1}$. (1) the dense case without noise (upper); (2) the dense case with noise (middle); (3) the sparse case with noise (lower).}
	\label{fig:simres_X1}
\end{figure}

We apply both DFPCA and FPCAder to $\Xp{1}$. We calculate the logarithm of ISE for the first two derivative-based eigenfunctions, $\widehat{\phi}_1^{[1]}$ and $\widehat{\phi}_2^{[1]}$, which together explain approximately 95\% of the total variance in most simulations. Similarly, the RE and logarithm of MSE are computed for the first two derivative-based eigenvalues (i.e., $\widehat{\nu}^{[1]}_1$ and $\widehat{\nu}^{[1]}_2$) and scores (i.e., $\widehat{\xi}^{[1]}_1$ and $\widehat{\xi}^{[1]}_2$), respectively. Figure~\ref{fig:simres_X1} presents the corresponding simulation results, which illustrates DFPCA performs better than FPCAder in all three settings. In particular, the errors are extremely small under the no noise setting for both methods, whereas the DFPCA still outperforms FPCAder, with the mean of RMISE 0.4\% vs 0.9\%. When data are noisy but dense, the mean of RMISE of DFPCA and FPCAder increase to 3\% and 5\%, respectively. When data are sparse and noisy, the DFPCA still performs better than FPCAder, with the mean of RMISE 12\% vs 14\%. We provide the simulation results for $\Xp{2}$ in Appendix~\ref{sec:appendixSimulation}, which also illustrates the DFPCA outperforms FPCAder. Moreover, an example of the estimation of the derivative of covariance is provided in Appendix~\ref{sec:appendixSimulation} to illustrate DFPCA is better than FPCAder. 

It is noted that DFPCA relies on the FACE algorithm with an additive penalty in P-splines. We find that the pre-defined parameters in Algorithm~\ref{alg:FACE_with_additive_penalty} have little impact on the DFPCA results. Firstly, the choice of the number of B-spline basis functions is not critical, as smoothness is primarily controlled by the penalty terms. We set $c=38$ in simulation to provide enough flexibility to capture the underlying trend. Secondly, the B-spline degree should satisfy $q \geq d+2$. We use cubic B-splines (i.e., $q=3$) which is a common choice in practice. At last, for the penalty orders, we consider three pairs $(l_1, l_2)=\left\{(1,2), (1,3), (2,3)\right\}$. For each pair, we apply the DFPCA method to $\Xp{1}$ under the dense and noisy setting. Table~\ref{tab:PO_selection} shows the corresponding DFPCA results for each pair, presented as mean (standard deviation). This suggests that the selection of penalty orders has little effect on the DFPCA performance, although the pair $(l_1, l_2)=(2,3)$ provides better overall results. 

\begin{table}[H]
	\centering
	\caption{DFPCA performance under different penalty orders $(l_1, l_2)=\left\{(1,2), (1,3), (2,3)\right\}$.}
	\label{tab:PO_selection}
	\begin{tabularx}{\textwidth}{c|X|X|X|X|X|X|X}
			\hline
			Penalty order & \multicolumn{2}{c|}{RE} & \multicolumn{2}{c|}{ISE} & \multicolumn{2}{c|}{MSE} &RMISE \\ 
			\cline{2-3}\cline{4-5}\cline{6-7}
			$(l_1, l_2)$& $k=1$ & $k=2$ &  $k=1$ & $k=2$ &  $k=1$ & $k=2$  &  \\ 
			\hline
			$(1,2)$ & 0.13 (0.03) & 0.11 (0.06)& 0.04 (0.01)& 0.06 (0.03)& 0.02 (0.00)&  0.10 (0.04)& 0.04 (0.01)\\ 
			$(1,3)$ & 0.13 (0.03) & 0.11 (0.06)& 0.04 (0.01)& 0.06 (0.03)& 0.02 (0.00)&  0.10 (0.04)& 0.04 (0.01)\\ 
			$(2,3)$ & 0.11 (0.03)& 0.08 (0.05)&0.03 (0.01)&0.05 (0.02)&0.02 (0.01)& 0.10 (0.03)& 0.03 (0.01)\\
			\hline
	\end{tabularx}
\end{table}

\subsection{Multivariate functional data}
\label{sec:simulation_multi}
We consider the multivariate functional data $X(\pointt)=(\Xp{1}(t), \Xp{2}(t))$ and apply DMFPCA based on Algorithm~\ref{alg:DMFPCA} under the three settings. To the best of our knowledge, there is no existing method for estimating derivative-based FPCA for multivariate functional data. Although it is possible to extend FPCAder to multivariate cases, it is unlikely to provide better estimation results than the DMFPCA, as we have shown that FPCAder is outperformed by DFPCA. 

For multivariate functional data, we evaluate the accuracy based on:
\begin{equation}
	\begin{aligned}
		\text{RE}(v_k, \widehat{v}_{k})&=\vert v_k-\widehat{v}_{k} \vert/v_k,\\
		\text{ISE}(\psi_{k}, \widehat{\psi}_{k})&=\sum_{p=1}^{P} \int \left(\psi_k^{[p]}(t_p)-\widehat{\psi}_k^{[p]}(t_p) \right)^2dt_p,\\
		\text{MSE}(\rho_k, \widehat{\rho}_k)&=\frac{ \sum_{i=1}^{n}(\rho_{ik}-\widehat{\rho}_{ik})^2/n}{\Var(\rho_k)},\\
		\text{RMISE}(\partial X, \widehat{\partial X})&=
		\frac{1}{P}\sum_{p=1}^{P}\frac{\sum_{i=1}^{n} \int (\partial\Xp{p}_i(t_p)-\widehat{\partial \Xp{p}_i}(t_p))^2dt_p}{\sum_{i=1}^{n} \int (\partial \Xp{p}_i(t_p))^2dt_p},
	\end{aligned}
\end{equation} 
where $v_k$, $\psi_k$ and $\rho_k$ are obtained by applying R function \texttt{MFPCA()} \citep{happ2018multivariate} to the true derivatives $\partial X$; $\widehat{v}_{k}$, $\widehat{\psi}_{k}$ and $\widehat{\rho}_k$ are estimated by applying DMFPCA based on Algorithm~\ref{alg:DMFPCA}. After estimating the DMFPCs $\widehat{\psi}_{k}$ and DMFPC-scores $\widehat{\rho}_k$, the multivariate Karhunen-Loève representation is used to construct the derivatives, i.e., $\widehat{\partial X}(\pointt)=\sum_{k=1}^{M} \widehat{\rho}_k \widehat{\psi}_k(\pointt)$.

\begin{figure}[H]
		\centering
		\includegraphics[width=\linewidth]{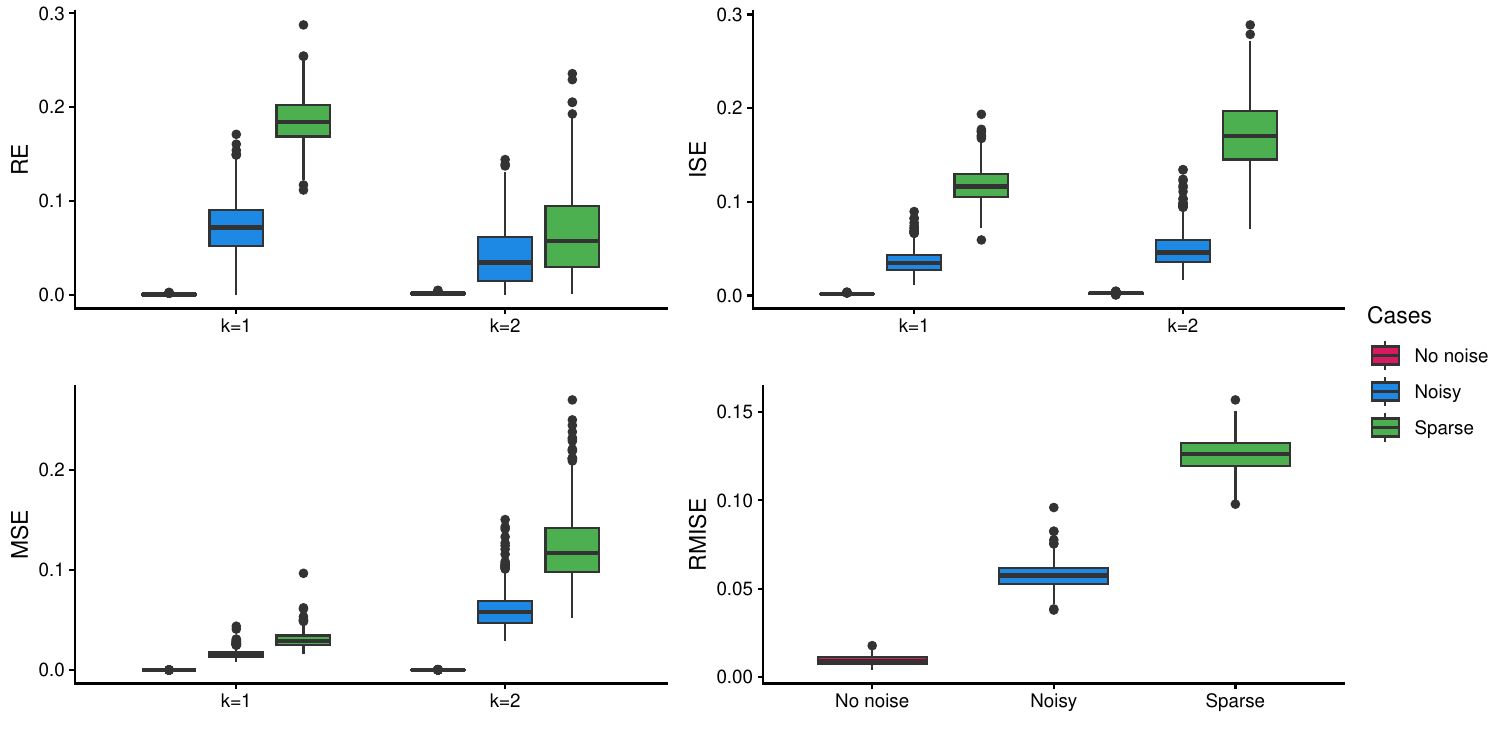}
		\caption{Simulation results for multivariate functional data under three settings.}
		\label{fig:simres}
\end{figure}

Figure~\ref{fig:simres} shows the RE, ISE, MSE and RMISE for the multivariate functional data under three settings (i.e., no noise, noisy, sparse and noisy). The RE of the first two eigenvalues are below 1\% under the no noise setting, and increase to approximately 10\% when data are noisy, and increase further when data are sparse. The ISE of DMFPCs are extremely small for the no noise case, with mean values around 0.002 for the first two DMFPCs. They increase to 0.036 for $k=1$ and 0.049 for $k=2$ when data are noisy, and 0.117 and 0.171 when data are sparse. The mean MSE of DMFPC-scores are less than 1\% for both $k=1$ and $k=2$ when data are without noise, and increase to 2\% for $k=1$ and 6\% for $k=2$ when data are noisy, to 3\% for $k=1$ and 12\% for $k=2$ when data are sparse. The mean RMISE of fitted derivatives are 1\% for the no noise case, 6\% for the noisy case and 13\% for the sparse case. 


%% file: main/application.tex
\section{Application} 
\label{sec:application}

Wearable sensors such as inertial measurement units (IMU) and electromyography (EMG) are widely used to capture human movement data. These data offer valuable information on gait patterns and movement abnormalities. Through quantifying variables such as angle, velocity and acceleration, joint kinematics analysis can improve sports performance, reduce injury risk and facilitate early detection of abnormal movement patterns. 

\begin{figure}[h]
	\centering
	\includegraphics[width=0.9\linewidth]{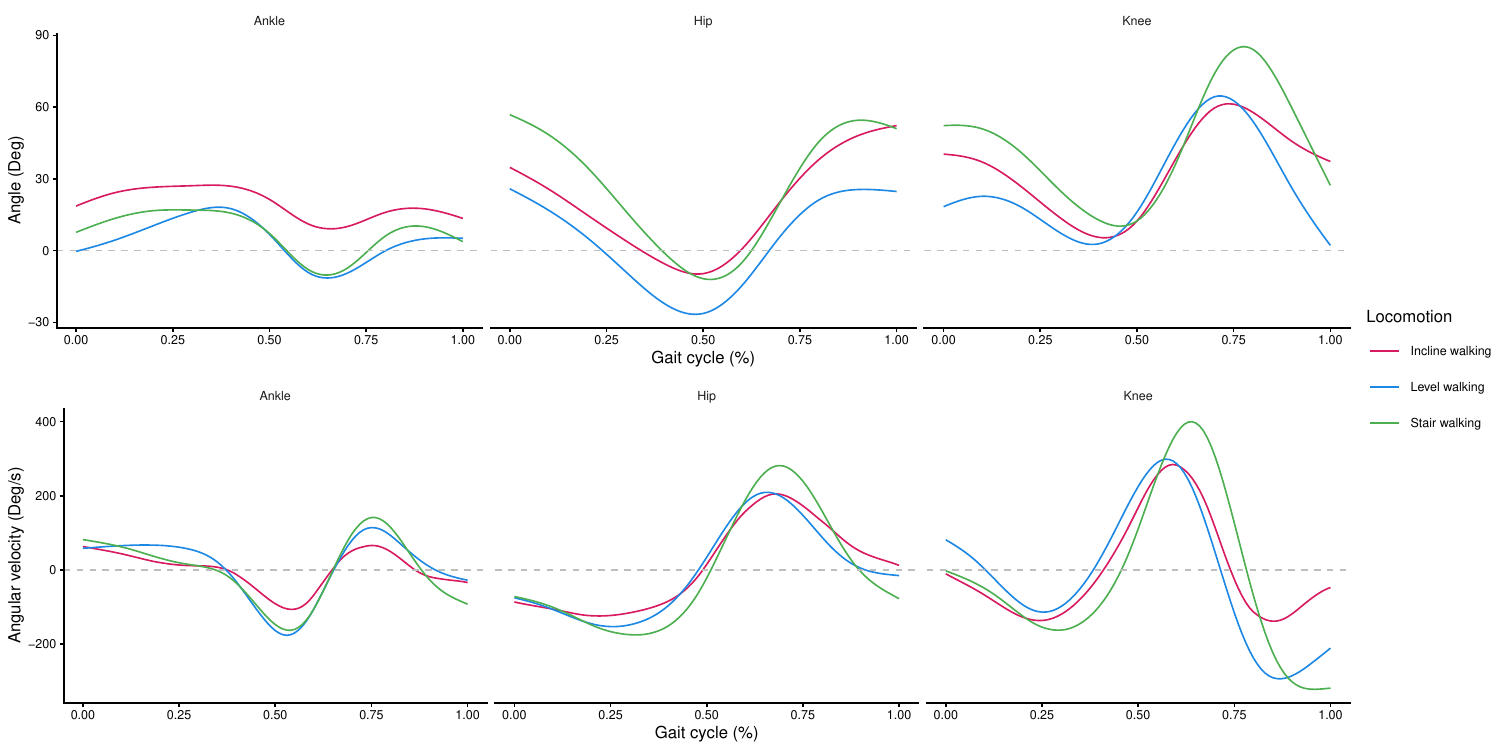}
	\caption{An example of ankle dorsiflexion, hip flexion and knee flexion for three participants (top panel) and their fitted angular velocities by DMFPCA (bottom panel).} 
	\label{fig:app_example}
\end{figure}

\cite{boo2025comprehensive} presented a dataset of human locomotion during daily activities collected from 120 male participants. Participants performed three locomotion tasks: level walking, incline walking and stair ascent. Reflective markers and surface EMG sensors were attached to the lower limb, such as right knee (RNKE), right ankle (RANK) and right hip (RASI), to measure joint kinematics and muscle activity, see Fig.2 of \cite{boo2025comprehensive}. We randomly selected 40 participants from the level, incline and stair walking groups, respectively. For each participant, we used data collected from the RANK, RASI and RNKE positions. The top panel of Figure~\ref{fig:app_example} shows an example of ankle dorsiflexion, hip flexion and knee flexion for three participants over a normalized gait cycle. Positive values represent dorsiflexion (ankle) or flexion (hip and knee), whereas negative values indicate plantarflexion or extension. The trajectories suggest a joint kinematic pattern. For instance, during a gait cycle in level walking, the transition to ankle plantarflexion around the mid-gait coincided with maximum hip extension and the onset of rapid knee flexion. On the other hand, joint angular velocity (i.e., the first derivative of joint angles) can measure the rate of change of joint angles over the gait cycle and characterize both the speed and timing of joint motion; however, it was not observed in the dataset. By applying DMFPCA to the joint angle trajectories, we aim to construct the joint angular velocities and to obtain their corresponding eigenfunctions and scores. 

\begin{figure}[H]
	\centering
	\includegraphics[width=0.8\linewidth]{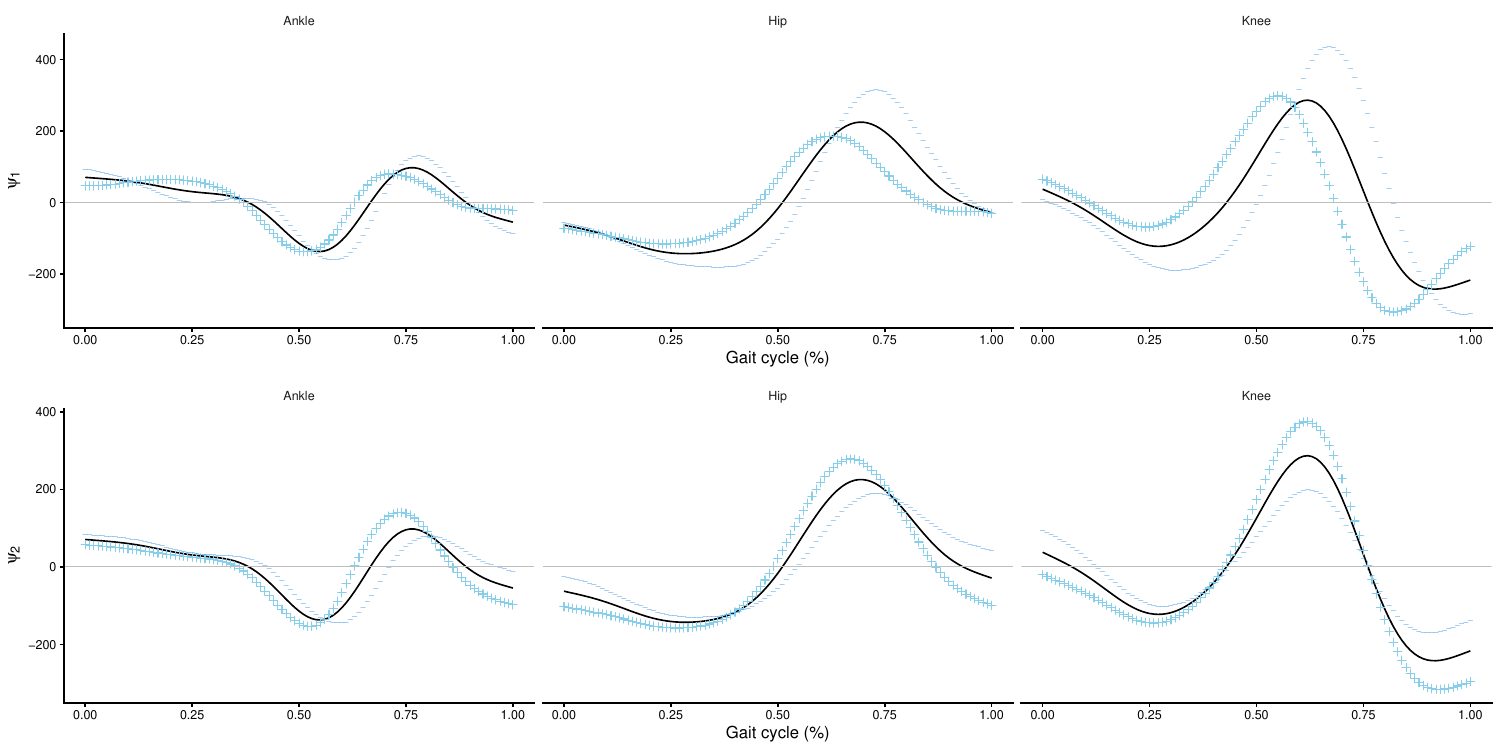}
	\caption{The mean functions (black) plus and minus the first two multivariate eigenfunctions of joint angular velocities (i.e., the first two DMFPCs of joint angle trajectories).} 
	\label{fig:app_eigfuns}
\end{figure}

Applying DMFPCA to joint angles retained the first six DMFPCs which together explained 96\% of the total variation. Figure~\ref{fig:app_eigfuns} presents the first two DMFPCs, which accounted for 60\% and 18\% of the total variance, respectively. Note that DMFPCs of joint angles correspond to multivariate eigenfunctions of joint angular velocities, and zero crossings for angular velocity reflect changes in movement direction. From Figure~\ref{fig:app_eigfuns}, participants with a higher first DMFPC-score experienced earlier zero crossings and smaller velocity magnitudes, and participants with a higher second DMFPC-score had earlier zero crossings and larger velocity magnitudes. This suggests the DMFPC-scores characterize key features of joint motion for each participant, such as the range of velocity and timing of directional change.

\begin{figure}[H]
	\centering
	\begin{subfigure}{0.45\textwidth}
		\centering
		\includegraphics[width=\linewidth]{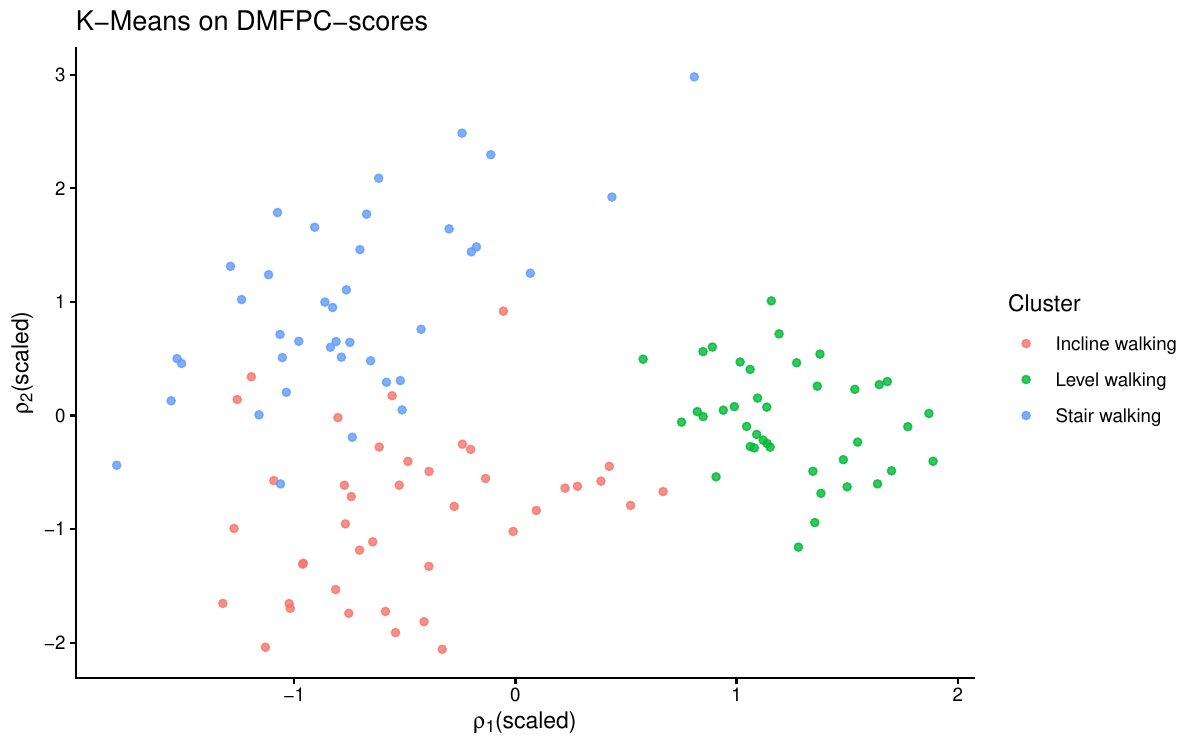}
	\end{subfigure}
	\begin{subfigure}{0.45\textwidth}
		\centering
		\includegraphics[width=\linewidth]{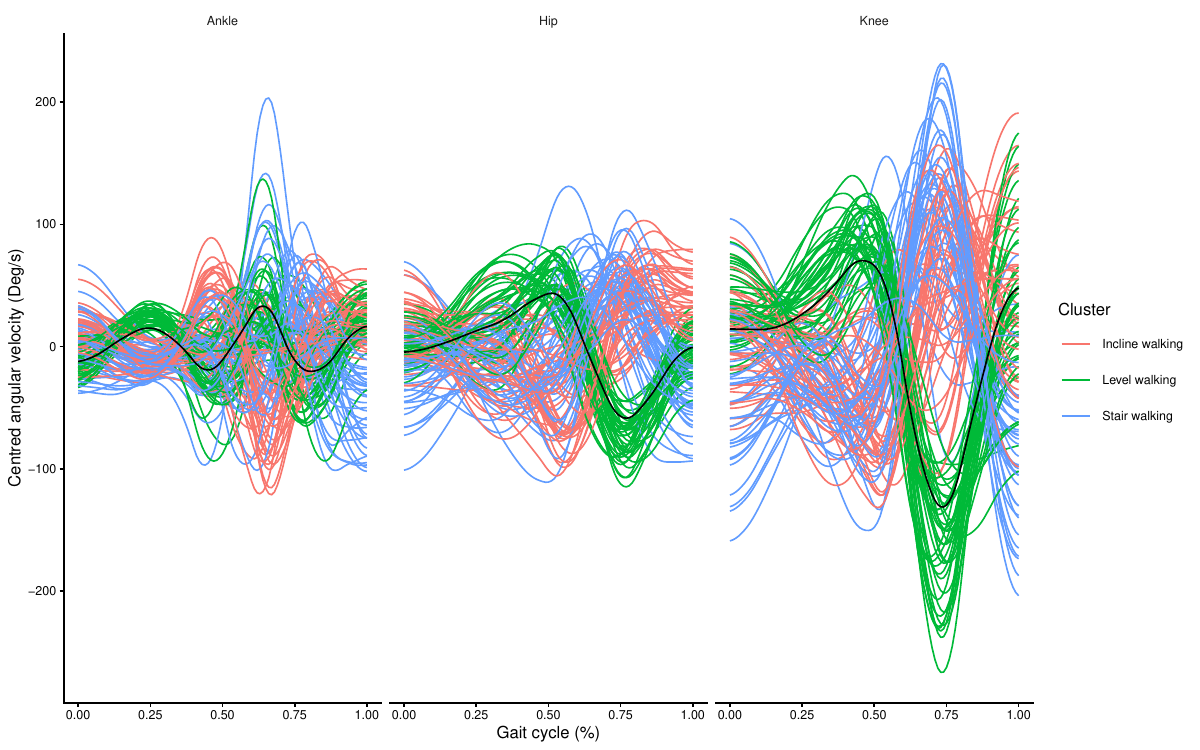}
	\end{subfigure}
	\caption{Results for K-means clustering using scaled DMFPC-scores. Left panel: the three clusters against the first two scaled DMFPC-scores. Right panel: fitted derivatives (after centering) together with the first DMFPC represented by the black line.}
	\label{fig:app_kmeans}
\end{figure}

We applied K-means clustering to the scaled DMFPC-scores and fixed the number of clusters to three based on the known locomotion tasks. Figure~\ref{fig:app_kmeans} presents the results for K-means clustering using DMFPC-scores. The left panel shows the three clusters against the first two scaled DMFPC-scores, which indicates that the first two DMFPC-scores clearly discriminate the three motions. More specifically, participants in the level walking group generally had higher first DMFPC-scores than those in the incline and stair walking groups, whereas participants in the stair walking group generally had higher second DMFPC-scores than those the incline walking group. Based on Figure~\ref{fig:app_eigfuns}, it further suggests that participants in the level walking group experienced an earlier time of directional change and a smaller velocity range compared to those in the incline and stair walking groups. It is evident that the level walking typically has a shorter stance phase and a smaller range of joint motion compared with incline and stair walking. Moreover, the right panel of Figure~\ref{fig:app_kmeans} illustrates the first DMFPC (represented in black) primarily determined the cluster of level walking, as it captured the overall pattern of the fitted derivative curves within this group (shown in green). 

Next, we evaluated classification performance using the agreement between the ground-truth cluster labels and labels obtained after K-means clustering. Note that the ground-truth labels were defined a \textit{priori} when 40 participants were selected from each locomotion task. Based on K-means clustering of the DMFPC-scores, the cluster sizes for level, stair and incline walking were 40, 39, and 41, respectively. More specifically, all participants in the level walking task were correctly identified, 3 participants in the stair walking task were misclassified as incline walking and 2 participants in the incline walking task were misclassified as stair walking. Overall, the classification achieved a high accuracy of 0.96 (95\% CI: 0.91-0.99). This demonstrates that the DMFPC-scores provided a compact and powerful summary of key features of joint motion, which enabled accurate classification of different locomotion tasks. Additionally, we conducted a sensitivity analysis by varying the number of participants assigned to each locomotion task. We applied DMFPCA to the new samples and performed K-means clustering of the DMFPC-scores. It was found that the classification accuracy remained high. For instance, when ground-truth labels changed to 30, 50 and 40 (or to 30, 40 and 50) for level, incline and stair walking tasks, respectively, the classification accuracy was 0.94 with 95\% CI: 0.88-0.98 (or 0.93 with 95\% CI: 0.87-0.97).

Finally, angular velocities were constructed after estimating DMFPCs and DMFPC-scores of joint angle trajectories. The bottom panel of Figure~\ref{fig:app_example} shows the fitted angular velocities of three participants, where the stair walking tended to exhibit larger velocities at the ankle, hip and knee than the level and incline walking. In general, understanding joint angular velocity across different locomotion tasks may help clinicians evaluate how individual joint movement trajectories deviate from tasks. More importantly, DMFPC-scores provided a powerful and compact summary of joint motion and could be used to predict future outcomes such as risk of falls or injuries, which may assist clinicians in early risk identification.


%% file: main/discussion.tex
\section{Discussion}
\label{sec:discussion}

FPCA and derivative-based FPCA provide a comprehensive view of functional variability, where FPCA captures variation in observed functional data and derivative-based FPCA characterises underlying dynamic behaviour. The covariance estimation plays a fundamental role in both FPCA and derivative-based FPCA. The FACE algorithm offers higher estimation accuracy and computationally efficiency in covariance estimation than local linear smoothers and bivariate P-splines \citep{xiao2016fast, cui2023fast}, where the smooth estimator is represented in a sandwich form. This motivates extending FACE to estimate the derivative of the covariance function. The current literature relies on local linear smoothers to estimate the derivative of the covariance \citep{liu2009estimating, dai2018derivative, grithFunctionalPrincipalComponent2018}. In this paper, we propose a new approach by incorporating the FACE algorithm with an additive penalty in P-splines. Including an extra penalty term in P-splines has been shown to improve estimates of derivatives in one-dimensional cases \citep{simpkin2013additive, hernandez2023derivative}. When estimating the derivatives of the covariance, the additional penalty enforces smoothness in both dimensions of the covariance surface, which prevents undersmoothing and leads to improved estimation accuracy. Moreover, the smooth estimator of the derivative of the covariance retains a sandwich form, therefore ensuring that the estimator is symmetric and positive semi-definite. 

Following the FACE algorithm with an additive penalty, we proceed to derivative-based FPCA. We propose a new method to estimate the derivative-based FPCs without directly requiring the eigendecomposition of the derivative of the covariance. We use a mixed model equation to estimate the derivative-based scores, which avoids calculating the inverse of the derivative of the covariance in BLUP. Based on this framework, we develop a DMFPCA method, which is the first method to the best of our knowledge to consider the derivative-based FPCA in multivariate cases.

There are several limitations of our study. Firstly, we note that the FACE with an additive penalty in P-splines is primarily designed for densely observed data, as shown in \cite{xiao2016fast} and \cite{cui2023fast} where the original FACE algorithm was proposed and applied under dense settings. Secondly, even if an additive penalty term is included, the boundary effects in P-splines may still lead to reduced accuracy for estimating the derivative-based FPCA. Thirdly, we have focused on the scenario where the first-order derivative is considered but our method could be used for higher-order cases.

In conclusion, this paper proposes the FACE algorithm with an additive penalty in P-splines and applies it to derivative-based FPCA. The derivative-based FPCA and its extension to multivariate cases help identify variations in dynamic changes and provide a low-dimensional summary of these variations for functional trajectories.


%% file: main/Appendix.tex
\appendix
\section{Appendix} 
\renewcommand{\theequation}{A.\arabic{equation}}
\setcounter{equation}{0}
\renewcommand{\thefigure}{A.\arabic{figure}}
\setcounter{figure}{0}

\subsection{Proof of Lemma~\ref{lem:smoothermatrix}}
\label{sec:Lemma}

To prove Lemma~\ref{lem:smoothermatrix}, we start from \eqref{eq:FACE_eigdec}
\begin{equation}
	\BTB{-1/2} \left(w\Penalty_1 + (1-w)\Penalty_2 \right)\BTB{-1/2} = \U \text{diag}(\s) \U^{\T}.
\end{equation}
Given that $\lambda_{+} =\lambda_1 +\lambda_2$ and $w=\lambda_1/ \lambda_{+}$, we have $\lambda_{+} \left(w\Penalty_1 + (1-w)\Penalty_2 \right) =  \lambda_1 \Penalty_1 + \lambda_2 \Penalty_2$. Multiplying both sides of \eqref{eq:FACE_eigdec} by $\lambda_{+}$ leads to
\begin{equation}
	\BTB{-1/2} ( \lambda_1 \Penalty_1 + \lambda_2 \Penalty_2) \BTB{-1/2} = \U \lambda_{+}\text{diag}(\s) \U^{\T}.
\end{equation}
Multiplying both sides of the above equation on the left by $\U^{\T}$ and on the right by $\U$ gives

\begin{equation}
	\U^{\T} \BTB{-1/2} ( \lambda_1 \Penalty_1 + \lambda_2 \Penalty_2) \BTB{-1/2}\U =\lambda_{+} \text{diag}(\s),
\end{equation}
which can be further constructed into	
\begin{equation}
	\label{eq:FACE_eigdec_step1}
	\U^{\T} \BTB{-1/2} ( \B^{\T} \B + \lambda_1 \Penalty_1 + \lambda_2 \Penalty_2 ) \BTB{-1/2}\U = \mathbf{I}_c + \lambda_{+} \text{diag}(\s).
\end{equation}
Taking the inverse of both sides of \eqref{eq:FACE_eigdec_step1} gives
\begin{equation}
	\label{eq:FACE_eigdec_step2}
	\U^{\T} \BTB{1/2} ( \B^{\T} \B + \lambda_1 \Penalty_1 + \lambda_2 \Penalty_2)^{-1}  \BTB{1/2}\U = \left\{\mathbf{I}_c + \lambda_{+} \text{diag}(\s)\right\}^{-1}.
\end{equation}
Define $\A_0=(\B^T \B)^{-1/2}\U$, $\A_s=\B \A_0$, $\Sigmas_s = \left\{\mathbf{I}_c + \lambda_{+} \text{diag} (\s)\right\}^{-1}$. By multiplying both sides of \eqref{eq:FACE_eigdec_step2} on the left by $\A_s$ and on the right by $\A_s^{\T}$, we have
\begin{equation}
	\label{eq:FACE_eigdec_step3}
	\B ( \B^{\T} \B + \lambda_1 \Penalty_1 + \lambda_2 \Penalty_2)^{-1}  \B^{\T} =\A_s \Sigmas_s \A_s^{\T},
\end{equation}
where the left-hand side is exactly the smoother matrix $\Smoother$. Therefore, we complete the proof of Lemma~\ref{lem:smoothermatrix}.


\subsection{Simulation}
\label{sec:appendixSimulation}

\begin{figure}[H]
	\centering
	\includegraphics[width=0.9\linewidth]{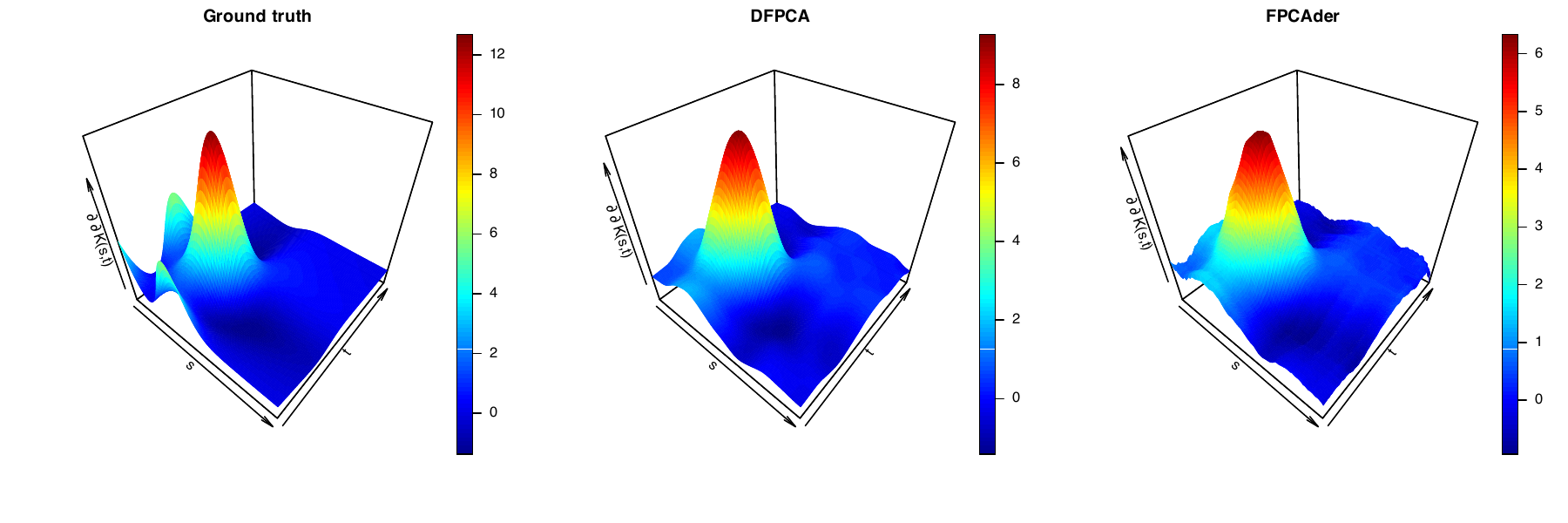}
	\caption{The derivative of covariance surface of $\partial \Xp{1}$ under the dense and noisy setting.} 
	\label{fig:covest}
\end{figure}

Section~\ref{sec:simulation_uni} provides the simulation results of $\Xp{1}$ in terms of the RE for eigenvalues, ISE for eigenfunctions, MSE for scores and RMISE for derivatives under three different settings.  In particular, Figure~\ref{fig:covest} presents the estimation result of the derivative of covariance surface for $\Xp{1}$ under the dense and noisy setting. The ground truth surface is computed from the given true derivative functions of $\Xp{1}$. The DFPCA and FPCAder methods estimate the surface based on dense but noisy observations of $\Xp{1}$. Although the DFPCA method loses accuracy near the boundaries due to the boundary effect of P-splines, it outperforms the FPCAder method which leads to undersmoothing and wiggly eigenfunctions.

\begin{figure}[H]
	\centering
	\includegraphics[width=\linewidth]{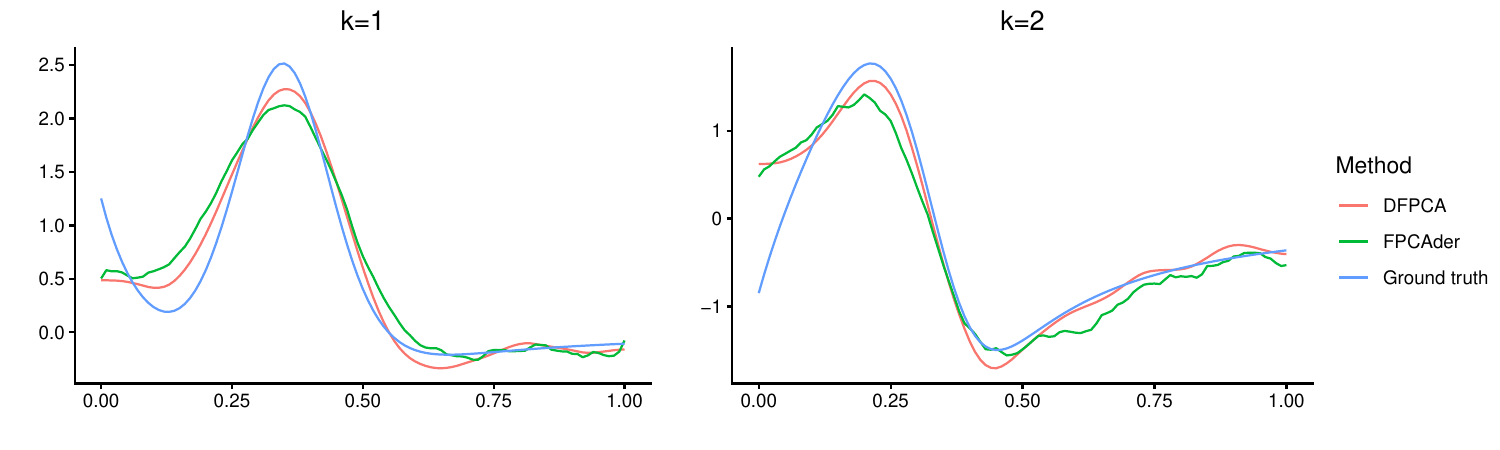}
	\caption{The derivative-based eigenfunctions of $\partial \Xp{1}$ under the dense and noisy setting.}
	\label{fig:eigfun}
\end{figure}

The first two eigenfunctions of the derivative of covariance surfaces (see Figure~\ref{fig:covest}) are provided in Figure~\ref{fig:eigfun}. We note that even if an additive penalty is included, the boundary effects in P-splines may still lead to reduced accuracy for derivative estimation, as illustrated by the differences between the blue (i.e., the ground truth) and red (i.e., DFPCA) lines. Nevertheless, the derivative-based eigenfunctions estimated by DFPCA still outperforms those estimated by FPCAder.

\begin{figure}[H]
	\centering
	\begin{subfigure}{\textwidth}
		\centering
		\includegraphics[width=\linewidth]{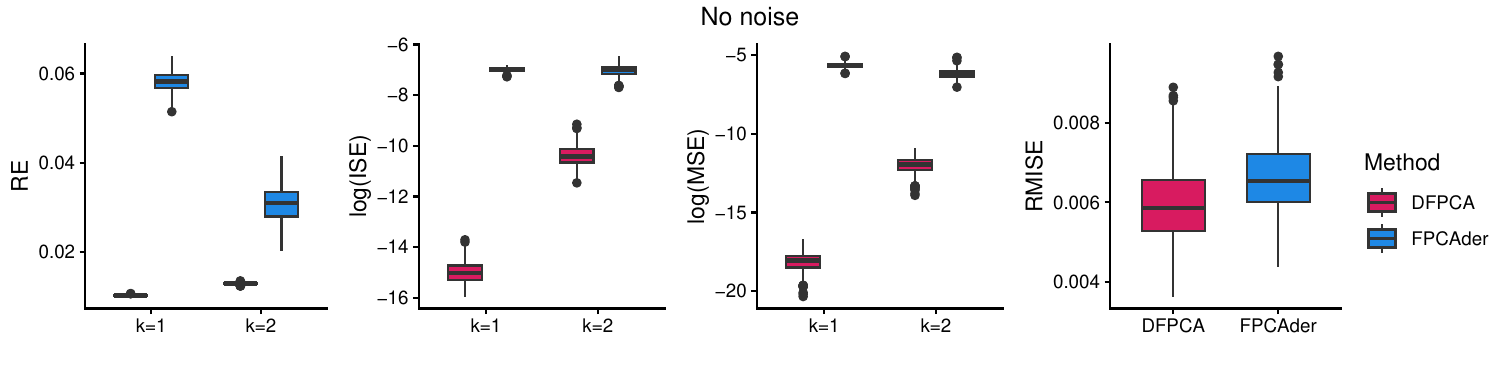}
	\end{subfigure}
	\\
	\begin{subfigure}{\textwidth}
		\centering
		\includegraphics[width=\linewidth]{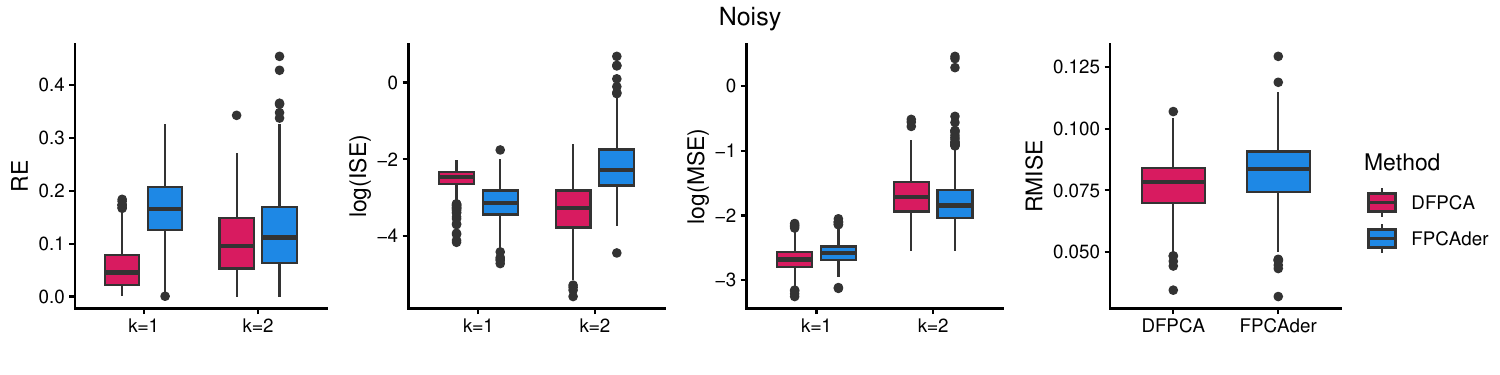}
	\end{subfigure}
	\\
	\begin{subfigure}{\textwidth}
		\centering
		\includegraphics[width=\linewidth]{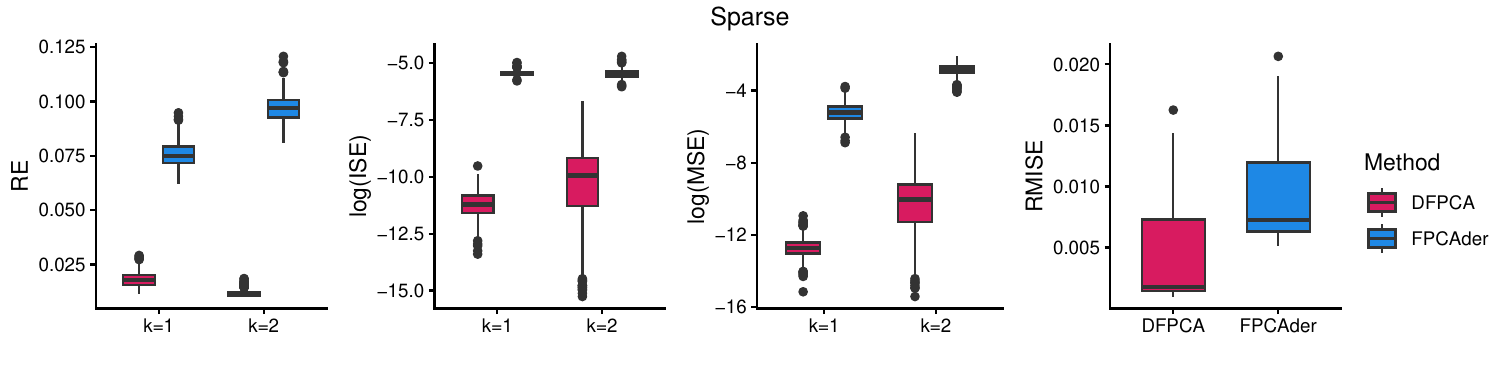}
	\end{subfigure}
	\caption{Simulation results for $\Xp{2}$, (1) the dense case without noise (upper); (2) the dense case with noise (middle); (3) the sparse case with noise (lower).}
	\label{fig:simres_X2}
\end{figure}

In the following, we present the simulation result of $\Xp{2}$. We apply both DFPCA and FPCAder to $\Xp{2}$. We calculate the logarithm of ISE for the first two derivative-based eigenfunctions, $\widehat{\phi}_1^{[2]}$ and $\widehat{\phi}_2^{[2]}$, which together explain approximately 95\% of the total variance in most simulations. Similarly, the RE and logarithm of MSE are computed for the first two derivative-based eigenvalues (i.e., $\widehat{\nu}^{[2]}_1$ and $\widehat{\nu}^{[2]}_2$) and scores (i.e., $\widehat{\xi}^{[2]}_1$ and $\widehat{\xi}^{[2]}_2$), respectively. Figure~\ref{fig:simres_X2} presents the corresponding simulation results, which illustrates DFPCA performs better than FPCAder in all three settings.

